\documentclass[twocolumn]{aastex631}
\usepackage{color}
\usepackage{tikz, graphicx}
\usepackage{xspace} 

\usepackage{url}
\usepackage{hyperref}
\usepackage{float}
\usepackage{soul}
\usepackage{bm}

\newcommand{\ixpe}{{IXPE}\xspace}
\newcommand{\integral}{{INTEGRAL}\xspace}
\newcommand{\maxi}{{MAXI}\xspace}
\newcommand{\nicer}{{NICER}\xspace}  

\newcommand\source{GS~1826$-$238\xspace}

\def\phabs{\texttt{phabs}}
\def\comptt{\texttt{comptt}}
\def\diskbb{\texttt{diskbb}}
\def\polconst{\texttt{polconst}}

\definecolor{Magenta}{cmyk}{0.1,0.8,0,0.1} 
\definecolor{Green}{rgb}{0.0, 0.5, 0.0}

\graphicspath{{./}{figures/}}

\submitjournal{ApJ}

\shorttitle{Observations of GS~1826$-$238 with IXPE}
\shortauthors{Capitanio et al.}

\begin{document}

\title{Polarization properties of the weakly magnetized neutron star X-ray binary GS 1826$-$238 \\ in the high soft state}

\author[0000-0002-6384-3027]{Fiamma Capitanio}
\affiliation{INAF Istituto di Astrofisica e Planetologia Spaziali, Via del Fosso del Cavaliere 100, 00133 Roma, Italy}
\author[0000-0002-6384-3023]{Sergio Fabiani}
\affiliation{INAF Istituto di Astrofisica e Planetologia Spaziali, Via del Fosso del Cavaliere 100, 00133 Roma, Italy}
\author[0000-0002-0642-1135]{Andrea Gnarini}
\affiliation{Dipartimento di Matematica e Fisica, Universit\`{a} degli Studi Roma Tre, Via della Vasca Navale 84, 00146 Roma, Italy}
\author[0000-0001-9442-7897]{Francesco Ursini}
\affiliation{Dipartimento di Matematica e Fisica, Universit\`{a} degli Studi Roma Tre, Via della Vasca Navale 84, 00146 Roma, Italy}
\author[0000-0003-1429-1059]{Carlo Ferrigno}
\affiliation{Department of Astronomy, University of Geneva, Ch. d'Ecogia 16, 1290, Versoix, Geneva, Switzerland}
\author[0000-0002-2152-0916]{Giorgio Matt}
\affiliation{Dipartimento di Matematica e Fisica, Universit\`{a} degli Studi Roma Tre, Via della Vasca Navale 84, 00146 Roma, Italy}
\author[0000-0002-0983-0049]{Juri Poutanen}
\affiliation{Department of Physics and Astronomy,  20014 University of Turku, Finland}
\affiliation{Space Research Institute of the Russian Academy of Sciences, Profsoyuznaya Str. 84/32, Moscow 117997, Russia}
\author[0000-0002-5817-3129]{Massimo Cocchi}
\affiliation{INAF -- Osservatorio Astronomico di Cagliari, via della Scienza 5, I-09047 Selargius (CA), Italy}
\author[0000-0001-7374-843X]{Romana Mikusincova}
\affiliation{Dipartimento di Matematica e Fisica, Universit\`{a} degli Studi Roma Tre, Via della Vasca Navale 84, 00146 Roma, Italy}
\author[0000-0003-2212-367X]{Ruben Farinelli}
\affiliation{INAF -- Osservatorio di Astrofisica e Scienza dello Spazio di Bologna, Via P. Gobetti 101, I-40129 Bologna, Italy}
\author[0000-0002-4622-4240]{Stefano Bianchi}
\affiliation{Dipartimento di Matematica e Fisica, Universit\`{a} degli Studi Roma Tre, Via della Vasca Navale 84, 00146 Roma, Italy}
\author[0000-0002-3010-8333]{Jari J. E. Kajava}
\affiliation{Department of Physics and Astronomy,  20014 University of Turku, Finland}
\affiliation{Aurora Technology for the European Space Agency (ESA), European Space Astronomy Centre (ESAC), Camino Bajo del Castillo s/n, 28692 Villanueva de la Cañada, Madrid, Spain}
\author[0000-0003-3331-3794]{Fabio Muleri}
\affiliation{INAF Istituto di Astrofisica e Planetologia Spaziali, Via del Fosso del Cavaliere 100, 00133 Roma, Italy}
\author[0000-0002-0778-6048]{Celia Sanchez-Fernandez}
\affiliation{ATG Europe for the European Space Agency (ESA), European Space Astronomy Centre (ESAC), Camino Bajo del Castillo s/n, 28692 Villanueva de la Cañada, Madrid, Spain}
\author[0000-0002-7781-4104]{Paolo Soffitta}
\affiliation{INAF Istituto di Astrofisica e Planetologia Spaziali, Via del Fosso del Cavaliere 100, 00133 Roma, Italy}
\author[0000-0002-7568-8765]{Kinwah Wu}
\affiliation{Mullard Space Science Laboratory, University College London, Holmbury St Mary, Dorking, Surrey RH5 6NT, UK}

\author[0000-0002-3777-6182]{Iv\'an Agudo}
\affiliation{Instituto de Astrof\'{i}sicade Andaluc\'{i}a -- CSIC, Glorieta de la Astronom\'{i}a s/n, 18008 Granada, Spain}
\author[0000-0002-5037-9034]{Lucio A. Antonelli}
\affiliation{INAF Osservatorio Astronomico di Roma, Via Frascati 33, 00040 Monte Porzio Catone (RM), Italy}
\affiliation{Space Science Data Center, Agenzia Spaziale Italiana, Via del Politecnico snc, 00133 Roma, Italy}
\author[0000-0002-4576-9337]{Matteo Bachetti}
\affiliation{INAF Osservatorio Astronomico di Cagliari, Via della Scienza 5, 09047 Selargius (CA), Italy}
\author[0000-0002-9785-7726]{Luca Baldini}
\affiliation{Istituto Nazionale di Fisica Nucleare, Sezione di Pisa, Largo B. Pontecorvo 3, 56127 Pisa, Italy}
\affiliation{Dipartimento di Fisica, Universit\`{a} di Pisa, Largo B. Pontecorvo 3, 56127 Pisa, Italy}
\author[0000-0002-5106-0463]{Wayne H. Baumgartner}
\affiliation{NASA Marshall Space Flight Center, Huntsville, AL 35812, USA}
\author[0000-0002-2469-7063]{Ronaldo Bellazzini}
\affiliation{Istituto Nazionale di Fisica Nucleare, Sezione di Pisa, Largo B. Pontecorvo 3, 56127 Pisa, Italy}
\author[0000-0002-0901-2097]{Stephen D. Bongiorno}
\affiliation{NASA Marshall Space Flight Center, Huntsville, AL 35812, USA}
\author[0000-0002-4264-1215]{Raffaella Bonino}
\affiliation{Istituto Nazionale di Fisica Nucleare, Sezione di Torino, Via Pietro Giuria 1, 10125 Torino, Italy}
\affiliation{Dipartimento di Fisica, Universit\`{a} degli Studi di Torino, Via Pietro Giuria 1, 10125 Torino, Italy}
\author[0000-0002-9460-1821]{Alessandro Brez}
\affiliation{Istituto Nazionale di Fisica Nucleare, Sezione di Pisa, Largo B. Pontecorvo 3, 56127 Pisa, Italy}
\author[0000-0002-8848-1392]{Niccol\`{o} Bucciantini}
\affiliation{INAF Osservatorio Astrofisico di Arcetri, Largo Enrico Fermi 5, 50125 Firenze, Italy}
\affiliation{Dipartimento di Fisica e Astronomia, Universit\`{a} degli Studi di Firenze, Via Sansone 1, 50019 Sesto Fiorentino (FI), Italy}
\affiliation{Istituto Nazionale di Fisica Nucleare, Sezione di Firenze, Via Sansone 1, 50019 Sesto Fiorentino (FI), Italy}
\author[0000-0003-1111-4292]{Simone Castellano}
\affiliation{Istituto Nazionale di Fisica Nucleare, Sezione di Pisa, Largo B. Pontecorvo 3, 56127 Pisa, Italy}
\author[0000-0001-7150-9638]{Elisabetta Cavazzuti}
\affiliation{Agenzia Spaziale Italiana, Via del Politecnico snc, 00133 Roma, Italy}
\author[0000-0002-0712-2479]{Stefano Ciprini}
\affiliation{Istituto Nazionale di Fisica Nucleare, Sezione di Roma ``Tor Vergata'', Via della Ricerca Scientifica 1, 00133 Roma, Italy}
\affiliation{Space Science Data Center, Agenzia Spaziale Italiana, Via del Politecnico snc, 00133 Roma, Italy}
\author[0000-0003-4925-8523]{Enrico Costa}
\affiliation{INAF Istituto di Astrofisica e Planetologia Spaziali, Via del Fosso del Cavaliere 100, 00133 Roma, Italy}
\author[0000-0001-5668-6863]{Alessandra De Rosa}
\affiliation{INAF Istituto di Astrofisica e Planetologia Spaziali, Via del Fosso del Cavaliere 100, 00133 Roma, Italy}
\author[0000-0002-3013-6334]{Ettore Del Monte}
\affiliation{INAF Istituto di Astrofisica e Planetologia Spaziali, Via del Fosso del Cavaliere 100, 00133 Roma, Italy}
\author[0000-0002-5614-5028]{Laura Di Gesu}
\affiliation{Agenzia Spaziale Italiana, Via del Politecnico snc, 00133 Roma, Italy}
\author[0000-0002-7574-1298]{Niccol\`{o} Di Lalla}
\affiliation{Department of Physics and Kavli Institute for Particle Astrophysics and Cosmology, Stanford University, Stanford, California 94305, USA}
\author[0000-0003-0331-3259]{Alessandro Di Marco}
\affiliation{INAF Istituto di Astrofisica e Planetologia Spaziali, Via del Fosso del Cavaliere 100, 00133 Roma, Italy}
\author[0000-0002-4700-4549]{Immacolata Donnarumma}
\affiliation{Agenzia Spaziale Italiana, Via del Politecnico snc, 00133 Roma, Italy}
\author[0000-0001-8162-1105]{Victor Doroshenko}
\affiliation{Institut f\"{u}r Astronomie und Astrophysik, Universit\"{a}t T\"{u}bingen, Sand 1, 72076 T\"{u}bingen, Germany}
\author[0000-0003-0079-1239]{Michal Dov\v{c}iak}
\affiliation{Astronomical Institute of the Czech Academy of Sciences, Bo\v{c}n\'{i} II 1401/1, 14100 Praha 4, Czech Republic}
\author[0000-0003-4420-2838]{Steven R. Ehlert}
\affiliation{NASA Marshall Space Flight Center, Huntsville, AL 35812, USA}
\author[0000-0003-1244-3100]{Teruaki Enoto}
\affiliation{RIKEN Cluster for Pioneering Research, 2-1 Hirosawa, Wako, Saitama 351-0198, Japan}
\author[0000-0001-6096-6710]{Yuri Evangelista}
\affiliation{INAF Istituto di Astrofisica e Planetologia Spaziali, Via del Fosso del Cavaliere 100, 00133 Roma, Italy}
\author[0000-0003-1074-8605]{Riccardo Ferrazzoli}
\affiliation{INAF Istituto di Astrofisica e Planetologia Spaziali, Via del Fosso del Cavaliere 100, 00133 Roma, Italy}
\author[0000-0003-3828-2448]{Javier A. Garcia}
\affiliation{California Institute of Technology, Pasadena, CA 91125, USA}
\author[0000-0002-5881-2445]{Shuichi Gunji}
\affiliation{Yamagata University,1-4-12 Kojirakawa-machi, Yamagata-shi 990-8560, Japan}
\author{Kiyoshi Hayashida}
\altaffiliation{Deceased}
\affiliation{Osaka University, 1-1 Yamadaoka, Suita, Osaka 565-0871, Japan}
\author[0000-0001-9739-367X]{Jeremy Heyl}
\affiliation{University of British Columbia, Vancouver, BC V6T 1Z4, Canada}
\author[0000-0002-0207-9010]{Wataru Iwakiri}
\affiliation{Department of Physics, Faculty of Science and Engineering, Chuo University, 1-13-27 Kasuga, Bunkyo-ku, Tokyo 112-8551, Japan}
\author[0000-0001-9522-5453]{Svetlana G. Jorstad}
\affiliation{Institute for Astrophysical Research, Boston University, 725 Commonwealth Avenue, Boston, MA 02215, USA}
\affiliation{Department of Astrophysics, St. Petersburg State University, Universitetsky pr. 28, Petrodvoretz, 198504 St. Petersburg, Russia}
\author[0000-0002-5760-0459]{Vladimir Karas}
\affiliation{Astronomical Institute of the Czech Academy of Sciences, Bo\v{c}n\'{i} II 1401/1, 14100 Praha 4, Czech Republic}
\author{Takao Kitaguchi}
\affiliation{RIKEN Cluster for Pioneering Research, 2-1 Hirosawa, Wako, Saitama 351-0198, Japan}
\author[0000-0002-0110-6136]{Jeffery J. Kolodziejczak}
\affiliation{NASA Marshall Space Flight Center, Huntsville, AL 35812, USA}
\author[0000-0002-1084-6507]{Henric Krawczynski}
\affiliation{Physics Department and McDonnell Center for the Space Sciences, Washington University in St. Louis, St. Louis, MO 63130, USA}
\author[0000-0001-8916-4156]{Fabio La Monaca}
\affiliation{INAF Istituto di Astrofisica e Planetologia Spaziali, Via del Fosso del Cavaliere 100, 00133 Roma, Italy}
\author[0000-0002-0984-1856]{Luca Latronico}
\affiliation{Istituto Nazionale di Fisica Nucleare, Sezione di Torino, Via Pietro Giuria 1, 10125 Torino, Italy}
\author[0000-0001-9200-4006]{Ioannis Liodakis}
\affiliation{Finnish Centre for Astronomy with ESO,  20014 University of Turku, Finland}
\author[0000-0002-0698-4421]{Simone Maldera}
\affiliation{Istituto Nazionale di Fisica Nucleare, Sezione di Torino, Via Pietro Giuria 1, 10125 Torino, Italy}
\author[0000-0002-0998-4953]{Alberto Manfreda}  
\affiliation{Istituto Nazionale di Fisica Nucleare, Sezione di Pisa, Largo B. Pontecorvo 3, 56127 Pisa, Italy}
\author[0000-0003-4952-0835]{Fr\'{e}d\'{e}ric Marin}
\affiliation{Observatoire Astronomique de Strasbourg, CNRS, Universit\'{e} de Strasbourg,  UMR 7550, 67000 Strasbourg, France}
\author[0000-0002-2055-4946]{Andrea Marinucci}
\affiliation{Agenzia Spaziale Italiana, Via del Politecnico snc, 00133 Roma, Italy}
\author[0000-0001-7396-3332]{Alan P. Marscher}
\affiliation{Institute for Astrophysical Research, Boston University, 725 Commonwealth Avenue, Boston, MA 02215, USA}
\author[0000-0002-6492-1293]{Herman L. Marshall}
\affiliation{MIT Kavli Institute for Astrophysics and Space Research, Massachusetts Institute of Technology, 77 Massachusetts Avenue, Cambridge, MA 02139, USA}
\author{Ikuyuki Mitsuishi}
\affiliation{Graduate School of Science, Division of Particle and Astrophysical Science, Nagoya University, Furo-cho, Chikusa-ku, Nagoya, Aichi 464-8602, Japan}
\author[0000-0001-7263-0296]{Tsunefumi Mizuno}
\affiliation{Hiroshima Astrophysical Science Center, Hiroshima University, 1-3-1 Kagamiyama, Higashi-Hiroshima, Hiroshima 739-8526, Japan}
\author[0000-0002-5847-2612]{C.-Y. Ng}
\affiliation{Department of Physics, University of Hong Kong, Pokfulam, Hong Kong}
\author[0000-0002-1868-8056]{Stephen L. O'Dell}
\affiliation{NASA Marshall Space Flight Center, Huntsville, AL 35812, USA}
\author[0000-0002-5448-7577]{Nicola Omodei}
\affiliation{Department of Physics and Kavli Institute for Particle Astrophysics and Cosmology, Stanford University, Stanford, California 94305, USA}
\author[0000-0001-6194-4601]{Chiara Oppedisano}
\affiliation{Istituto Nazionale di Fisica Nucleare, Sezione di Torino, Via Pietro Giuria 1, 10125 Torino, Italy}
\author[0000-0001-6289-7413]{Alessandro Papitto}
\affiliation{INAF Osservatorio Astronomico di Roma, Via Frascati 33, 00040 Monte Porzio Catone (RM), Italy}
\author[0000-0002-7481-5259]{George G. Pavlov}
\affiliation{Department of Astronomy and Astrophysics, Pennsylvania State University, University Park, PA 16801, USA}
\author[0000-0001-6292-1911]{Abel L. Peirson}
\affiliation{Department of Physics and Kavli Institute for Particle Astrophysics and Cosmology, Stanford University, Stanford, California 94305, USA}
\author[0000-0003-3613-4409]{Matteo Perri}
\affiliation{Space Science Data Center, Agenzia Spaziale Italiana, Via del Politecnico snc, 00133 Roma, Italy}
\affiliation{INAF Osservatorio Astronomico di Roma, Via Frascati 33, 00040 Monte Porzio Catone (RM), Italy}
\author[0000-0003-1790-8018]{Melissa Pesce-Rollins}
\affiliation{Istituto Nazionale di Fisica Nucleare, Sezione di Pisa, Largo B. Pontecorvo 3, 56127 Pisa, Italy}
\author[0000-0001-6061-3480]{Pierre-Olivier Petrucci}
\affiliation{Universit\'{e} Grenoble Alpes, CNRS, IPAG, 38000 Grenoble, France}
\author[0000-0001-7397-8091]{Maura Pilia}
\affiliation{INAF Osservatorio Astronomico di Cagliari, Via della Scienza 5, 09047 Selargius (CA), Italy}
\author[0000-0001-5902-3731]{Andrea Possenti}
\affiliation{INAF Osservatorio Astronomico di Cagliari, Via della Scienza 5, 09047 Selargius (CA), Italy}
\author[0000-0002-2734-7835]{Simonetta Puccetti}
\affiliation{Space Science Data Center, Agenzia Spaziale Italiana, Via del Politecnico snc, 00133 Roma, Italy}
\author[0000-0003-1548-1524]{Brian D. Ramsey}
\affiliation{NASA Marshall Space Flight Center, Huntsville, AL 35812, USA}
\author[0000-0002-9774-0560]{John Rankin}
\affiliation{INAF Istituto di Astrofisica e Planetologia Spaziali, Via del Fosso del Cavaliere 100, 00133 Roma, Italy}
\author[0000-0003-0411-4243]{Ajay Ratheesh}
\affiliation{INAF Istituto di Astrofisica e Planetologia Spaziali, Via del Fosso del Cavaliere 100, 00133 Roma, Italy}
\author[0000-0001-6711-3286]{Roger W. Romani}
\affiliation{Department of Physics and Kavli Institute for Particle Astrophysics and Cosmology, Stanford University, Stanford, California 94305, USA}
\author[0000-0001-5676-6214]{Carmelo Sgr\`{o}}
\affiliation{Istituto Nazionale di Fisica Nucleare, Sezione di Pisa, Largo B. Pontecorvo 3, 56127 Pisa, Italy}
\author[0000-0002-6986-6756]{Patrick Slane}
\affiliation{Center for Astrophysics, Harvard \& Smithsonian, 60 Garden St, Cambridge, MA 02138, USA}
\author[0000-0003-0802-3453]{Gloria Spandre}
\affiliation{Istituto Nazionale di Fisica Nucleare, Sezione di Pisa, Largo B. Pontecorvo 3, 56127 Pisa, Italy}
\author[0000-0002-8801-6263]{Toru Tamagawa}
\affiliation{RIKEN Cluster for Pioneering Research, 2-1 Hirosawa, Wako, Saitama 351-0198, Japan}
\author[0000-0003-0256-0995]{Fabrizio Tavecchio}
\affiliation{INAF Osservatorio Astronomico di Brera, via E. Bianchi 46, 23807 Merate (LC), Italy}
\author[0000-0002-1768-618X]{Roberto Taverna}
\affiliation{Dipartimento di Fisica e Astronomia, Universit\`{a} degli Studi di Padova, Via Marzolo 8, 35131 Padova, Italy}
\author{Yuzuru Tawara}
\affiliation{Graduate School of Science, Division of Particle and Astrophysical Science, Nagoya University, Furo-cho, Chikusa-ku, Nagoya, Aichi 464-8602, Japan}
\author[0000-0002-9443-6774]{Allyn F. Tennant}
\affiliation{NASA Marshall Space Flight Center, Huntsville, AL 35812, USA}
\author[0000-0003-0411-4606]{Nicholas E. Thomas}
\affiliation{NASA Marshall Space Flight Center, Huntsville, AL 35812, USA}
\author[0000-0002-6562-8654]{Francesco Tombesi}
\affiliation{Dipartimento di Fisica, Universit\`{a} degli Studi di Roma ``Tor Vergata'', Via della Ricerca Scientifica 1, 00133 Roma, Italy}
\affiliation{Istituto Nazionale di Fisica Nucleare, Sezione di Roma ``Tor Vergata'', Via della Ricerca Scientifica 1, 00133 Roma, Italy}
\affiliation{Department of Astronomy, University of Maryland, College Park, Maryland 20742, USA}
 
\author[0000-0002-3180-6002]{Alessio Trois}
\affiliation{INAF Osservatorio Astronomico di Cagliari, Via della Scienza 5, 09047 Selargius (CA), Italy}
\author[0000-0002-9679-0793]{Sergey S. Tsygankov}
\affiliation{Department of Physics and Astronomy,  20014 University of Turku, Finland}
\affiliation{Space Research Institute of the Russian Academy of Sciences, Profsoyuznaya Str. 84/32, Moscow 117997, Russia}
\author[0000-0003-3977-8760]{Roberto Turolla}
\affiliation{Dipartimento di Fisica e Astronomia, Universit\`{a} degli Studi di Padova, Via Marzolo 8, 35131 Padova, Italy}
\affiliation{Mullard Space Science Laboratory, University College London, Holmbury St Mary, Dorking, Surrey RH5 6NT, UK}
\author[0000-0002-4708-4219]{Jacco Vink}
\affiliation{Anton Pannekoek Institute for Astronomy \& GRAPPA, University of Amsterdam, Science Park 904, 1098 XH Amsterdam, The Netherlands}
\author[0000-0002-5270-4240]{Martin C. Weisskopf}
\affiliation{NASA Marshall Space Flight Center, Huntsville, AL 35812, USA}
\author[0000-0002-0105-5826]{Fei Xie}
\affiliation{Guangxi Key Laboratory for Relativistic Astrophysics, School of Physical Science and Technology, Guangxi University, Nanning 530004, China}
\affiliation{INAF Istituto di Astrofisica e Planetologia Spaziali, Via del Fosso del Cavaliere 100, 00133 Roma, Italy}
\author[0000-0001-5326-880X]{Silvia Zane}
\affiliation{Mullard Space Science Laboratory, University College London, Holmbury St Mary, Dorking, Surrey RH5 6NT, UK}


\correspondingauthor{Fiamma Capitanio}
\email{fiamma.capitanio@inaf.it}

\begin{abstract}

The launch of the \textit{Imaging X-ray Polarimetry Explorer} (\ixpe) on 2021 December 9 has opened a new window in X-ray astronomy. 
We report here the results of the first IXPE observation of a weakly magnetized neutron star, GS~1826$-$238, performed on 2022 March 29--31 when the source was in a high soft state. An upper limit (99.73$\%$ confidence level) of 1.3\% for the linear polarization degree is obtained over the \ixpe 2--8 keV energy range. Coordinated \integral and \nicer observations were carried out simultaneously with \ixpe. The spectral parameters obtained from the fits to the broad-band spectrum were used as inputs for Monte Carlo simulations considering different possible geometries of the X-ray emitting region. Comparing the \ixpe upper limit with these simulations,  we can put constraints on the geometry and inclination angle of \source.
\end{abstract}



\section{Introduction} 
\label{sec:intro}

Weakly magnetized neutron stars in low-mass X-ray binaries (NS-LMXBs) are believed to accrete via Roche-lobe overflow from a stellar companion, which is typically a main sequence star with a mass lower than $\sim 1 \, M_{\odot}$ or an evolved white dwarf. 
These objects are highly variable in the X-rays at the timescale ranging from milliseconds to years. The classification of NS-LMXBs is historically based on the tracks that they draw on the so called color-color diagram  \citep[CCD,][]{Hasinger,vanderKlis95}. The sources are divided as a function of the X-ray luminosity as follows: a) high soft state (HSS) Z-sources ($>10^{38}$~erg\,s$^{-1}$); b) HSS bright atoll sources ($10^{37}-10^{38}$~erg\,s$^{-1}$); c) low hard state (LHS) atoll sources ($\sim 10^{36}$~erg\,s$^{-1}$) \citep[][and references therein]{vdKlis2006}.
The ``Z'' and ``atoll'' terms directly derive from the shape of the track in the CCDs.
The majority of persistent NS-LMXB are generally observed either in HSS or (less frequently) in LHS, but most  of the transients and several persistent  sources can perform state transitions from LHS to HSS and vice-versa in a relatively short timescale \citep{vdKlis2006}. 

The emission of this class of sources consists of two main spectral components: a soft ($<1$ keV) thermal component, produced by a relatively cold, optically thick, accretion disk, and a hard component, that can be modeled with Comptonization in a hot, relatively optically thin, electron plasma (often called {\it corona}) \citep{Done07}. Moreover, the frequent observation of an iron emission line at $\sim$ 6--7 keV, especially in the HSS sources \citep{Ludlam}, is likely a  signature of reflection by a colder medium (such as the geometrically thin accretion disk itself). In addition, the HSS spectra could show transient hard tails detected well beyond the Comptonized component, and up to $\sim$ 200--300 keV  whose origin is unclear \citep[see][and references therein]{Paizis06}. 
In LHS (but rarely also in HSS) NS-LMXBs typically show X-ray bursts, which are occasional powerful flashes (with their fluence of $\sim 10^{40}$~erg on a $\sim 100$~s timescale) due to a thermonuclear runaway in the dense H+He layer at the neutron star surface \citep{Lewin}. 
The evolution of the physical parameters  (plasma temperature, accretion rate, inner disk radius, etc.) defines the characteristics of the spectral states. For example, LHS plasma is much hotter and transparent (electron temperature $kT >20$ keV, Thomson optical depth $\tau \sim 2$) with respect to the HSS ones ($kT \sim 3$ keV, $\tau > 5$ depending on the geometry of the plasma itself). 

The presence of the NS surface stops the accretion flow forming a transition layer  between the disk and the NS surface. This layer is also named  spreading (SL) or boundary (BL) layer. In particular, the BL is the part of the accretion disk where the gas decelerates, while the SL is the gas layer at the NS surface, which can extend to high latitudes \citep{Inogamov99,suleimanov2006}.
In one of the most accredited model, the Eastern model \citep{Mitsuda84}, the soft component originates in the accretion disk, while the electron corona comptonizes the seed photons emitted by the NS surface and/or the boundary/spreading layer. Recently, \citet{Long} published a significant detection of a polarization signal (in the energy range 4--8 keV) in Sco X-1 with the {PolarLight} \citep{Feng} instrument. Their results, and in particular
the polarization angle roughly aligned with the radio jet, favor an electron corona located in the spreading/transition layer.
Timing analysis of these sources also supports the presence of the spreading layer, which may be even directly responsible of the emission of the hard component \citep{Gilfanov03,Revnivtsev06}. As discussed in \citet{Revnivtsev13}, on the base of the {RXTE} data, the hard component of NS-LMXB spectra can be modeled with a diluted blackbody. 
However, high sensitivity spectroscopy together with broad spectral coverage, such those permitted by {BeppoSAX} or {NuSTAR}, have shown that the hard emission is compatible with a comptonization spectrum \citep[see, e.g.,][and references therein]{Iaria,DiSalvo} rather than a diluted blackbody. Therefore, the nature of the hard component in the NS-LMXB spectra still remains an unresolved issue.  
In this framework, spectroscopy cannot help because of degeneracy in the parameter space providing information on the shape and extension of the region where Comptonization occurs. 
Polarimetry is the key to identify the nature and the geometry of the system removing degeneracy left by spectroscopy. In fact, different geometries and viewing angles result in quite different polarization degree (PD) and polarization angle (PA).
 


\subsection{\source}

\source is an accreting NS-LMXB. Until 2016 it was classified as an atoll source in the hard spectral state. The peculiarity of this source was the presence of extremely regular X-ray bursts over a range of several years \citep{Cocchi00,Zamfir12}. For this reason it is also known as ``clocked burster''. The clocked bursts occurred when \source was in the hard state as indicated by the CCD \citep{Cocchi11,Sanchez}, while during the occasional short transitions to the HSS (happened before MJD~57500), the bursts occurred  less regularly and were often shorter than in the hard state \citep{Chenevez16}.  At the beginning of 2016, \source underwent a major transition to the HSS. Since then, the source remained in the same state until the observational campaign described in this paper. The characteristics of the \source binary system are poorly known. As reported by \citet{Homer98}, a low amplitude modulation present in the optical light curve and the lack of eclipses imply a probable inclination of less than 70\degr. Other authors report tighter constraints. For example, \citet{Johnston20} modelled  multi-epoch X-ray bursts from \source with Markov chain Monte Carlo (MCMC) simulations obtaining an inclination angle of  $i\sim69^{+2}_{-3}$ deg. \citet{Mescheryakov11} estimated an inclination angle of $62.5\pm5.5$  deg from the mean optical flux and the amplitude of periodic modulations in the optical light curve.

\begin{figure*}
\centering
\includegraphics[angle=0,scale=0.36]{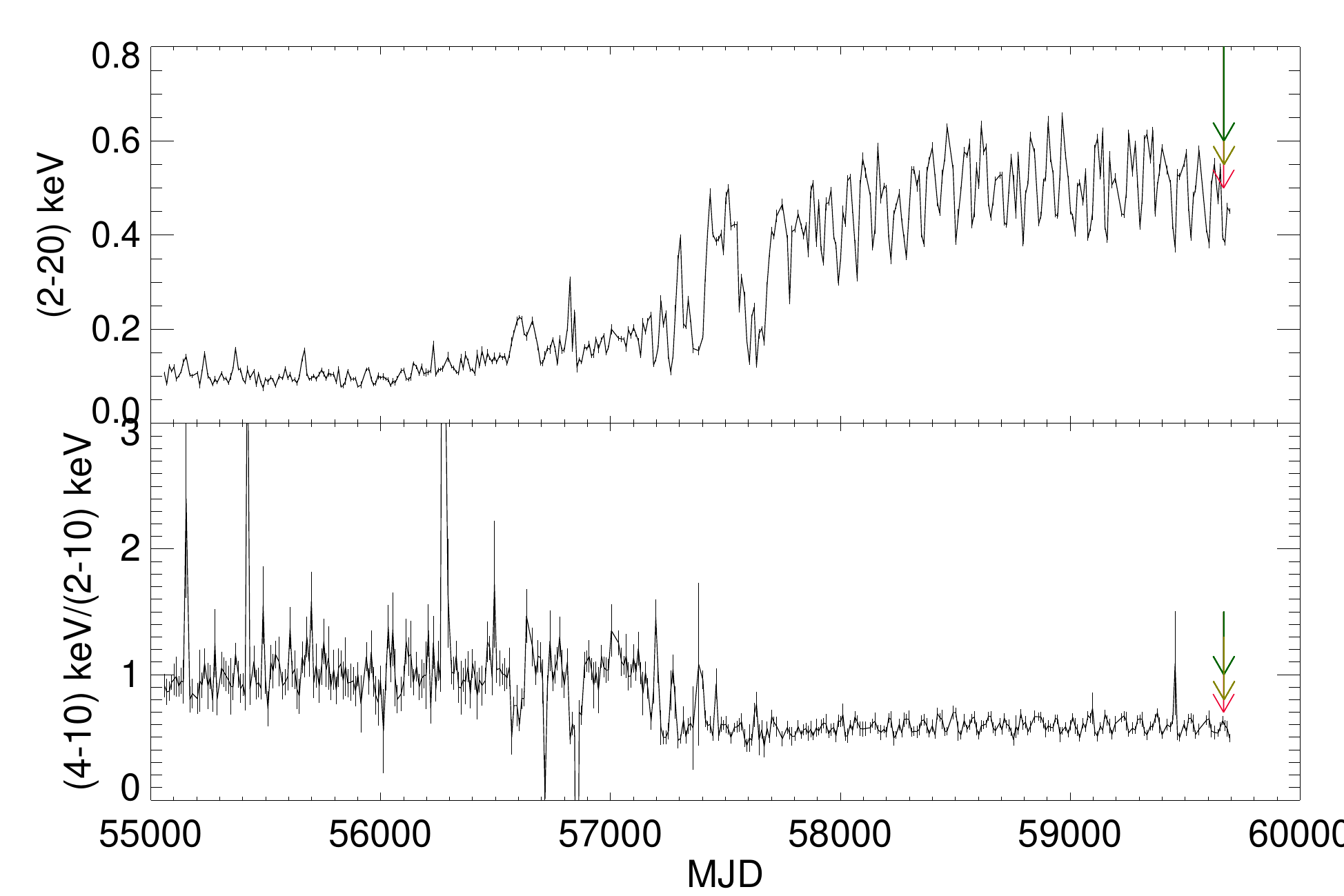}
\includegraphics[angle=0,scale=0.35]{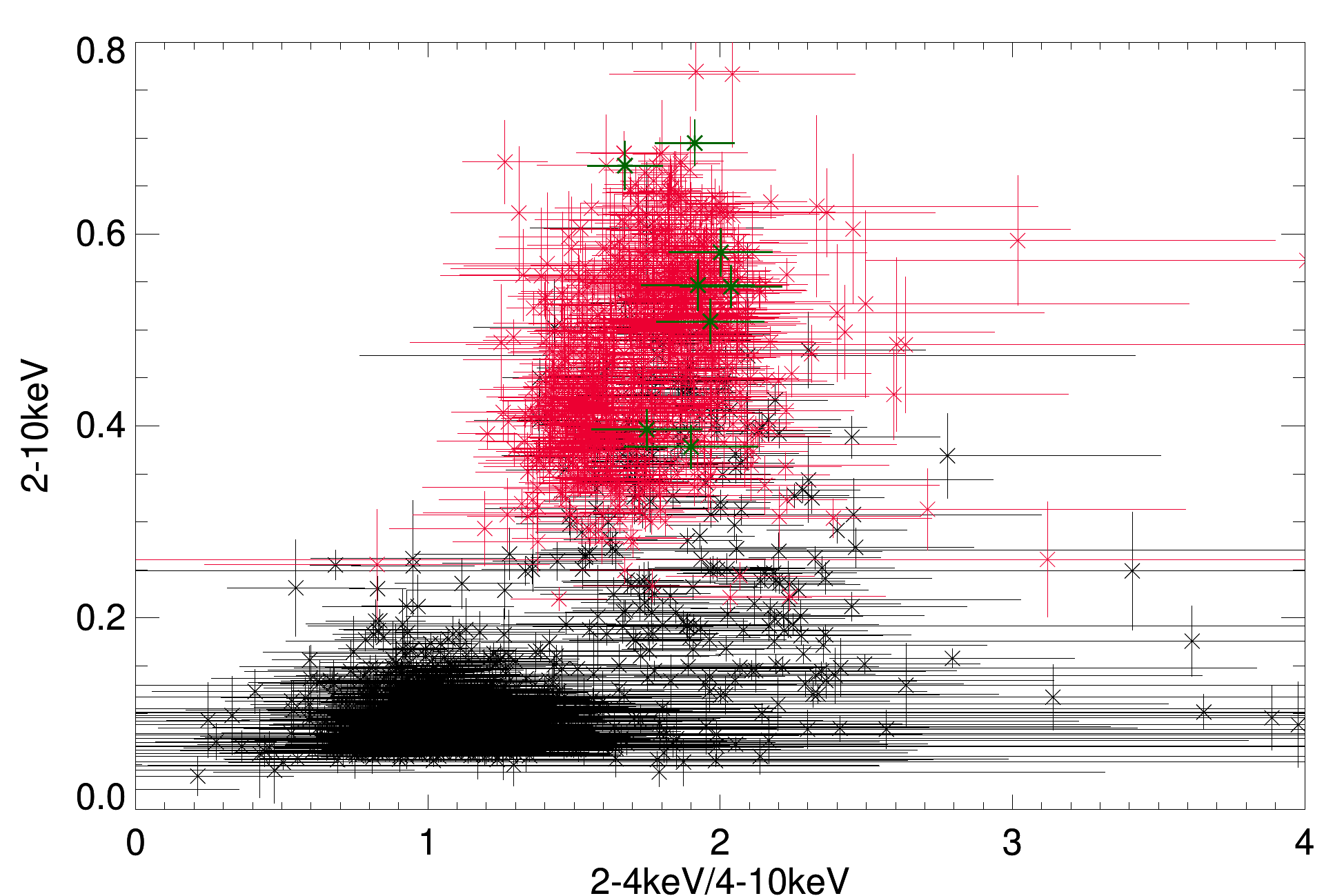}
\caption{Left top panel: the 2--20 keV \maxi light curve of \source  in units of ph\,cm$^{-2}$s$^{-1}$. Left bottom panel: Hardness-ratio (2--4~keV/4--10~keV) as a function of time. The green, gray and red arrows indicate the \ixpe, \nicer and \integral observation dates, respectively.
Right panel: the hardness-intensity diagram (HID) of \source derived from the \maxi data. The red points correspond to the HSS after the major transition in MJD~57500  (the black points represent the HID before MJD~57500), while the green points correspond to the times of the \nicer, \ixpe and \integral/JEM-X observations reported in this paper.  Flux variations among the green points are mostly due to the \maxi spurious modulation \citep{Mihara}. }
\label{fig:gs1826_longB}
\end{figure*}

\section{Data Reduction and analysis}

\subsection{The long time behavior of \source}

The left-top panel of Figure~\ref{fig:gs1826_longB} reports the 2--10~keV \maxi \citep{Matsuoka09} light curve of \source, while the left-bottom panel shows the hardness ratio (HR). 
The major transition of \source{} to the HSS is clearly visible in the \maxi light curve and in the HR at about MJD~57500. 
After that date, the large and periodic ($P\sim$72~d) flux variations correspond  to only slight variations in the HR (left panels of Figure~\ref{fig:gs1826_longB}) probably due to   a spurious 72-day oscillation sometimes present in the \maxi light curves~\citep{Mihara}. The right panel of Figure~\ref{fig:gs1826_longB}  shows the  hardness--intensity diagram (HID) for the sources based on the \maxi data.\footnote{We report here the HID and not the CCD for the  \source~\maxi data, because the errors in the CCD are too large to obtain a clear diagram.} 
The red points represent the values of the HID after the major transition to HSS and are all concentrated in a narrow range of intensity and hardness. This implies that, after MJD~57500, there were no transitions back to LHS. On the contrary, the spreading of the black points is due to several short transitions to the HSS before MJD~57500. 

\ixpe observed the source on 2022 March 29--31. 
A coordinated observational campaign with \nicer and \integral was performed simultaneously with \ixpe. The dates and the duration of the observations are reported in Table~\ref{tab:obs}. 
 An X-ray burst was detected in the JEM-X data in a time period not overlapping with \nicer and \ixpe observations (MJD~59667). The science window containing the X-ray burst (id:248500190010) was excluded from data analysis.
The green points in Figure~\ref{fig:gs1826_longB} represent the values of HID at the time of the \ixpe, \nicer and \integral observations.

\begin{deluxetable}{cccc}
\tablecaption{
\ixpe, \nicer and \integral observation log \label{tab:obs} }
\tablecolumns{4}
\tablewidth{1pt}
\tablehead{
\colhead{Telescope} &
\colhead{Obsid} &
\colhead{Date}  &
\colhead{Net exposure (ks)} 
}
\startdata
\ixpe & {01002801} & {2022-03-29/31}  & 92 \\
\nicer  & {5050310103} & {2022-03-30} & 6.4\\
\integral  & {2485/1970005} & {2022-03-28/30} & 139/108\tablenotemark{a} \\
\enddata
\tablenotetext{a}{JEM-X1/JEM-X2 exposure time.}
\end{deluxetable}

\subsection{\ixpe data}

The \textit{Imaging X-ray Polarimetry Explorer} \citep[\ixpe,][]{Weisskopf2022} is a NASA/ASI mission launched on 2021 December 9. \ixpe is observing all major classes of galactic and extragalactic X-ray sources, 
providing space, energy and time resolved polarimetry \citep{Soffitta}.
With respect to the previous X-ray polarimetric mission, {OSO}-8, \ixpe\ needs about two orders of magnitude less exposure time to reach the same sensitivity, and it provides imaging capability with $\leq30\arcsec$ angular resolution over $>11\arcmin$ field of view, together with  1--2\,$\mu$s timing accuracy and a moderate spectral resolution typical for proportional counters.
It  consists of three X-ray telescopes with identical mirror modules and identical polarization-sensitive imaging detector units (DUs) at their focus. 
The \ixpe observation took place on 2022 March 29--31, for a total net exposure time of 85~ks after taking into account Earth occultations. 

The \ixpe data extraction was performed by means of the \ixpe collaboration software tool \textsc{ixpeobssim} \citep{Baldini2022} version 26.3.2: \texttt{xppicorr} to apply the energy calibration with in-flight calibration sources (as such an correction was not implemented yet in the official pipeline at the time of the observation),  \texttt{xpselect} to filter data and \texttt{xpbin} to apply different binning algorithms for generating images and spectra. Rebinning and spectro-polarimetric analysis was performed with {\sc ftools} and \textsc{xspec} (HEASOFT version 6.30.1). We compared the results of the polarimetric analysis obtained with both \textsc{xspec} and \textsc{ixpeobssim} tools (\texttt{pcube}). While \textsc{xspec} requires the definition of a spectro-polarimetric model, \textsc{ixpeobssim} allows a model independent analysis that computes the polarization only on the basis of detected photons. 
The \textsc{ixpeobssim} response matrices version v010 were employed, corresponding to the latest available version in the HEASARC database. 
Data analysis was performed following the unweighted method.\footnote{In the unweighted analysis method, equal weights are assigned to each photo-electron track, regardless of its shape.} 

The statistical uncertainties of  PD and PA when using \textsc{ixpeobssim} are calculated with the assumption that the Stokes parameters are normally distributed and uncorrelated, and that PD and PA are considered independent, as described in \cite{Kislat2015}. 
We report these uncertainties in the tables as 68.27\% (1-$\sigma$) confidence level.
The uncertainties from the \textsc{xspec} analysis reported in the tables are computed with the \texttt{error} command of \textsc{xspec} for one parameter of interest.


It is worth noting that the PD and PA are, actually, not independent. The contours representing the 68.27\%,  95.45\% and  99.73\% confidence levels of the joint measurement of the PD and PA are a more appropriate method to represent the uncertainties. 
With \textsc{ixpeobssim} such contours are derived as described in \citet{Weisskopf2010}, \citet{Strohmayer2013} and  \citet{Muleri2022} by using the parameters obtained by the \texttt{pcube} algorithm itself. 
In the \textsc{xspec} the contours are obtained using the \texttt{steppar} command for two parameters of interest.
The upper limits to the PD are based upon its error in one dimension, without regard to the value of the PA. Therefore, they are computed using a $\chi^2$ with one degree of freedom.

Source and background regions where selected from the image of each DU. The source is centered in a circular region of 60\arcsec\ in radius. The background is extracted from an annular region with the internal and external radii of 180\arcsec\  and 240\arcsec, respectively. The background is almost negligible with respect to the source. The ratio of counts of background over the source (by scaling for the extraction region areas) is only $\sim 0.3\%$. 

\subsubsection{The \ixpe~ spectrum}

The \ixpe~ light curve and HR are substantially constant so we extracted the nine \ixpe~ Stokes parameters ($I$, $Q$ and $U$ for each DU) integrating over the entire observation. However, it should be noted that they were not compatible with the \nicer+JEMX spectra due to an improper correction of telescope vignetting, caused by the off axis pointing of \source{ } still present at the date of the observation. Due to \source{ } brightness the systematic effect induced is highly significant in the energy spectrum.\footnote{ https://heasarc.gsfc.nasa.gov/FTP/ixpe/data/obs/01/01002801/README}
It must be remarked, however, that this problem affects in the same way $I$, $Q$ and $U$, and therefore the PD and PA are not affected.

\subsection{\nicer data}

\nicer performed four observations of the source, with continuous exposure, in the period 2022 March 28--31. During the first two observations significant variability in the HR did not permit to extract a single averaged spectrum. For this reason we used in the joint fit only the third observation, ObsID 5050310103, that was simultaneous with \ixpe and has an exposure time of 6.4 ks. The \nicer data were reduced using \textsc{heasoft} 6.30 and the \textsc{nicerl2} task to apply standard calibration and screenings, with \textsc{caldb} version 20210707.

\subsection{\integral data}

\integral observed the source from 2022-03-28 17:25 to 2022-03-30 23:40 UT for a total of 186 ks. \integral data were reduced using the latest release of the standard On-line Scientific Analysis (OSA, version 11.2), distributed by the INTEGRAL Science Data Centre \citep[ISDC,][]{Courvoisier03} through the multi-messenger online data analysis platform \citep[MMODA,][]{Neronov21}. This target of opportunity observations were performed using hexagonal dithering to maintain \source in the fully coded field of view of JEM-X, the \integral X-ray telescope \citep{Lund03}. The JEM-X spectra were extracted in the range 3--35 keV with a response matrix with 16 standard channels. A systematic error of 1.5\% was added in quadrature for the spectral analysis. Even if the \integral observation did not overlap exactly the \ixpe and \nicer observations, the JEM-X spectrum  was in good agreement with the \nicer one. Because the JEM-X HR did not change significantly during the observation, it was possible to extract the averaged spectrum. Only the JEM-X data were used for the spectral extraction because IBIS, the $\gamma$-ray energy detector \citep{Ubertini99,Lebrun03}, did not detect the source with a 3-$\sigma$ upper limit on the flux of $\sim10^{-11}\,\mathrm{erg\,cm^{-2}\,s^{-1}}$ (3\,mCrab) in the 28--40 keV energy range, implying that the high-energy tail was not present.

\begin{figure} 
\centering
\vspace{+0.5cm}
\hspace{-1cm}
\includegraphics[angle=-90,scale=0.35]{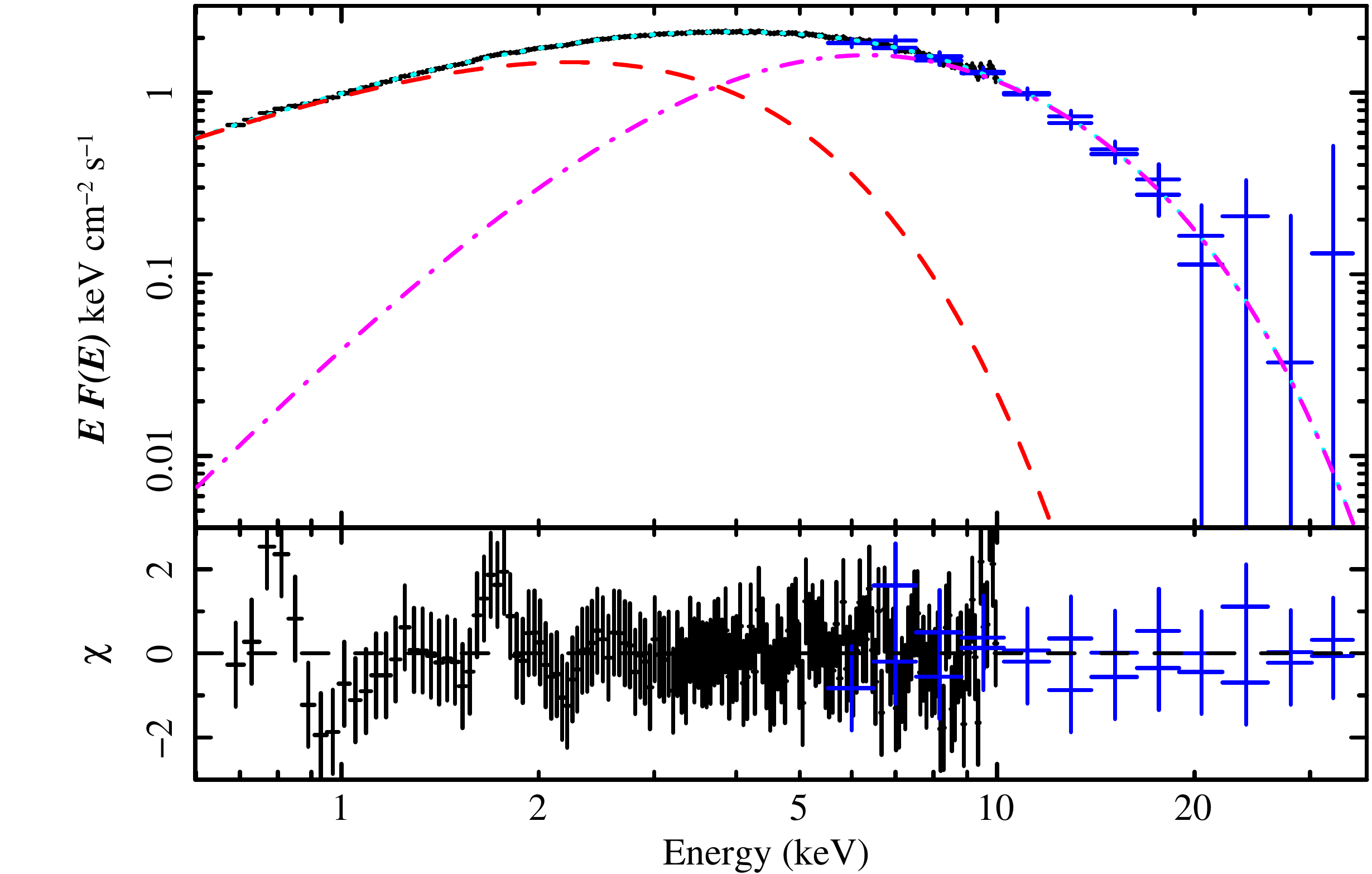}
\caption{An unfolded spectrum of \source as observed by \nicer\ (black crosses) and  JEM-X1 and 2 (both shown as blue crosses). The red dashed line represents the accretion disk emission (\diskbb\ in xspec); the pink dot--dashed line represent the emission of the Comptonized component (\comptt\ in xspec) The spectral parameters are given in Table~\ref{tab:fitpar}.  
The residuals below 2 keV are due to the \nicer instrumental issue as reported by \citet{Miller18} and do not affect significantly the continuous spectrum (see text for details). }
\label{eeuf_fig}
\end{figure}

\begin{deluxetable}{ccc}
\tablecaption{
Best-fit spectral parameters  obtained from the \nicer and JEM-X data}
\label{tab:fitpar}
\tablecolumns{6}
\tablewidth{0pt}
\tablehead{
\colhead{Fit parameter} &
\colhead{Slab/Sphere} 
}
\startdata
$N_{\rm H} ($10$^{22}$ cm$^{-2}$)\tablenotemark{a} & 0.351$^{+0.004}_{-0.005}$\\ 
$kT_{\rm in}$ (keV)\tablenotemark{b}                     & 0.94$\pm$0.1 \\ 
$N_{\rm d} (R_{\rm in}$  km)\tablenotemark{c}                     &  277$^{+134}_{-55}$(14$^{+3}_{-2}$) \\ 
$kT_{0}$  (keV)\tablenotemark{d}                      &1.3$\pm$0.2       \\
$kT_{\rm e}$  (keV)\tablenotemark{e}                      &2.7$^{+3.0}_{-0.2}$ \\
$\tau_{0,\rm slab}$($\tau_{0,\rm sphere}$)\tablenotemark{f}                        &4.9$^{+1.8}_{-3.2}$ (10.8$^{+3.5}_{-6.8}$) \\
$N_{\rm C}$\tablenotemark{g}                              & 0.3$\pm$0.1 \\ 
 $\chi_{\rm red}^{2}$ (d.o.f.)\tablenotemark{h} & 0.7 (172) \\ 
 $f_{(2-8\,\rm keV)}$ (erg cm$^{-2}$ s$^{-1}$)\tablenotemark{i} & 4.42$\times$10$^{-9}$ \\
  $f_{(2-4\,\rm keV)}$ (erg cm$^{-2}$ s$^{-1}$)\tablenotemark{i} &2.24$\times$10$^{-9}$ \\
  $f_{(4-8\,\rm keV)}$ (erg cm$^{-2}$ s$^{-1}$)\tablenotemark{i} &2.18$\times$10$^{-9}$ \\
  $f_{\rm disk}^{\rm ph}$ / $f_{\rm tot}^{\rm ph}$\tablenotemark{j}    & 0.55 \\
 $f_{\rm disk}^{\rm ene}$/ $f_{\rm tot}^{\rm ene}$\tablenotemark{k}    & 0.45 \\
\enddata
\tablecomments{Both slab and sphere geometries give identical spectral parameters except for the value of plasma optical depth. }
\tablenotetext{a}{Equivalent hydrogen column density.} 
\tablenotetext{b}{Inner disk temperature.}
\tablenotetext{c}{\diskbb\ normalization parameter $N_{\rm  d}=(R_{\rm  in}/10~\mbox{kpc})^2\cos\theta$, where $R_{\rm in}$ is the disk inner radius in km  and $\theta$  is the viewing angle ($\theta$=60$\degr$).}
\tablenotetext{d}{Seed photons temperature.} 
\tablenotetext{e}{Electron temperature.} 
\tablenotetext{f}{Plasma optical depths for spherical and slab geometry.} 
\tablenotetext{g}{Normalization of the \comptt\ component.} 
\tablenotetext{h}{Reduced $\chi^{2}$ and the degrees of freedom.} 
\tablenotetext{i}{The unabsorbed flux in the energy range specified by the subscript.} 
\tablenotetext{j}{Fraction of all photons in the  2--8 keV range in the \diskbb\ component.}
\tablenotetext{k}{Fraction of the energy flux in the  2--8 keV in the \diskbb\ component.} 
\end{deluxetable}

\section{Results}\label{results}

\subsection{Spectroscopy of \source}

We carried out the spectral analysis of the joint \textit{NICER} and JEM-X spectrum using \textsc{xspec}, version 12.12.1. The model used for the fitting procedure is a disk black-body component \citep{Mitsuda84} plus a Comptonization of soft photons in a hot plasma \citep{Titarchuk94}. Both components are modified by interstellar absorption. The \textsc{xspec} syntax of the model has the form: {\phabs*(\diskbb+\comptt)}. No reflection component and iron line are needed in the spectral fit. We performed the spectral fitting for two different geometries: slab and sphere. The spectral parameters obtained from the fitting procedures are reported in Table~\ref{tab:fitpar}. The  corresponding unfolded spectrum is shown in Figure~\ref{eeuf_fig}. The features present in the residuals are due to \nicer instrumental issue \citep{Strohmayer18, Miller18}. We  verified, in two different ways, that these features do not affect the continuous spectrum: 1) modelling the features by adding two Gaussian line profiles to the model; 2) ignoring the \nicer spectrum in the range  0--2.3 keV. In both cases the spectral parameters remain consistent within the errors. 

The spectrum of \source is consistent with those reported in literature for a weakly-magnetized NS-LMXB in HSS, with the low temperature ($\sim$2.7 keV) and a highly opaque electron plasma \citep[see, for example,][]{Paizis06}. As expected, both geometries are consistent with the data (see Table~\ref{tab:fitpar} for details).

\subsection{Polarization measurements}
\label{pol}

\begin{figure*}[ht!]
\includegraphics[angle=0,scale=0.16]{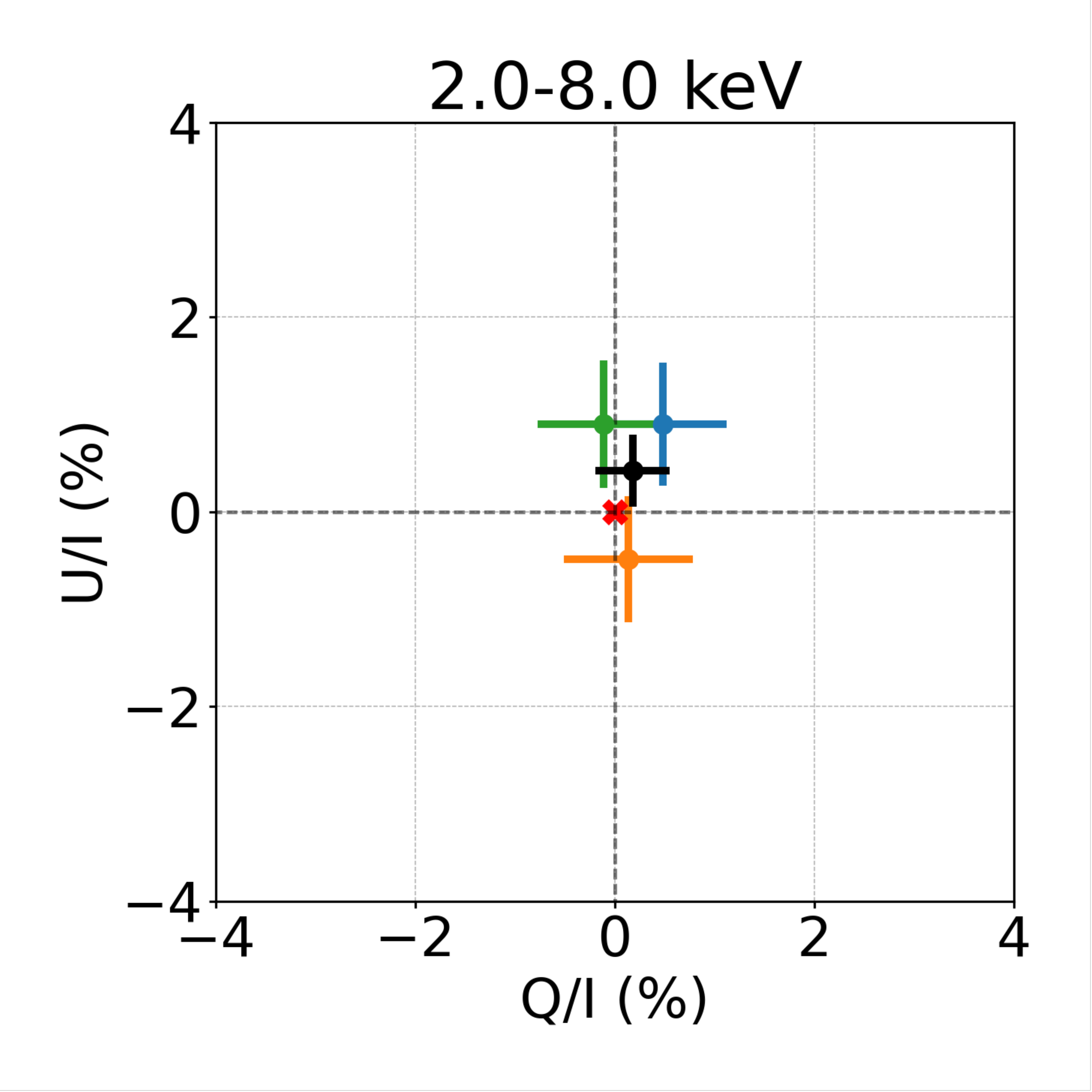}
\includegraphics[angle=0,scale=0.16]{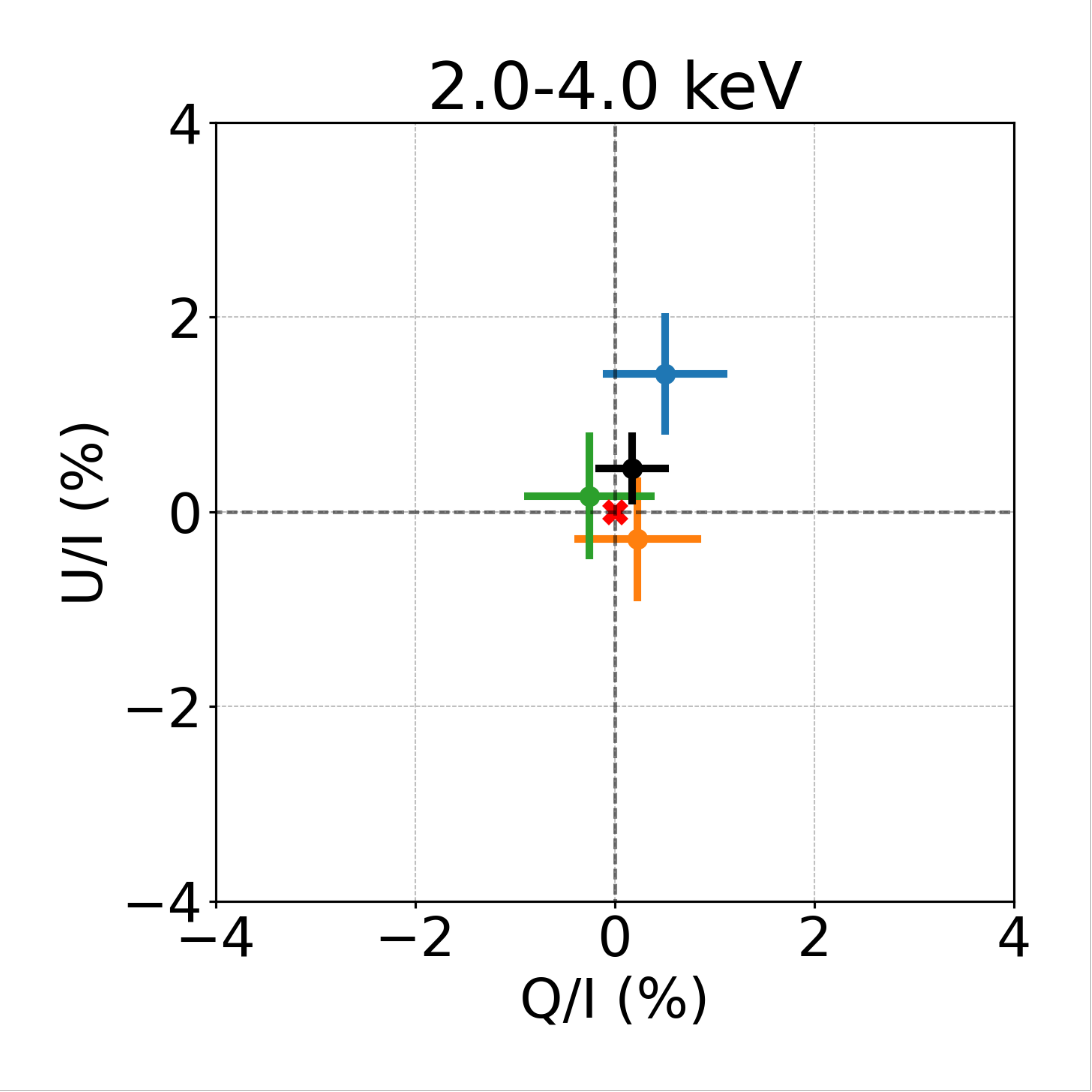}
\includegraphics[angle=0,scale=0.16]{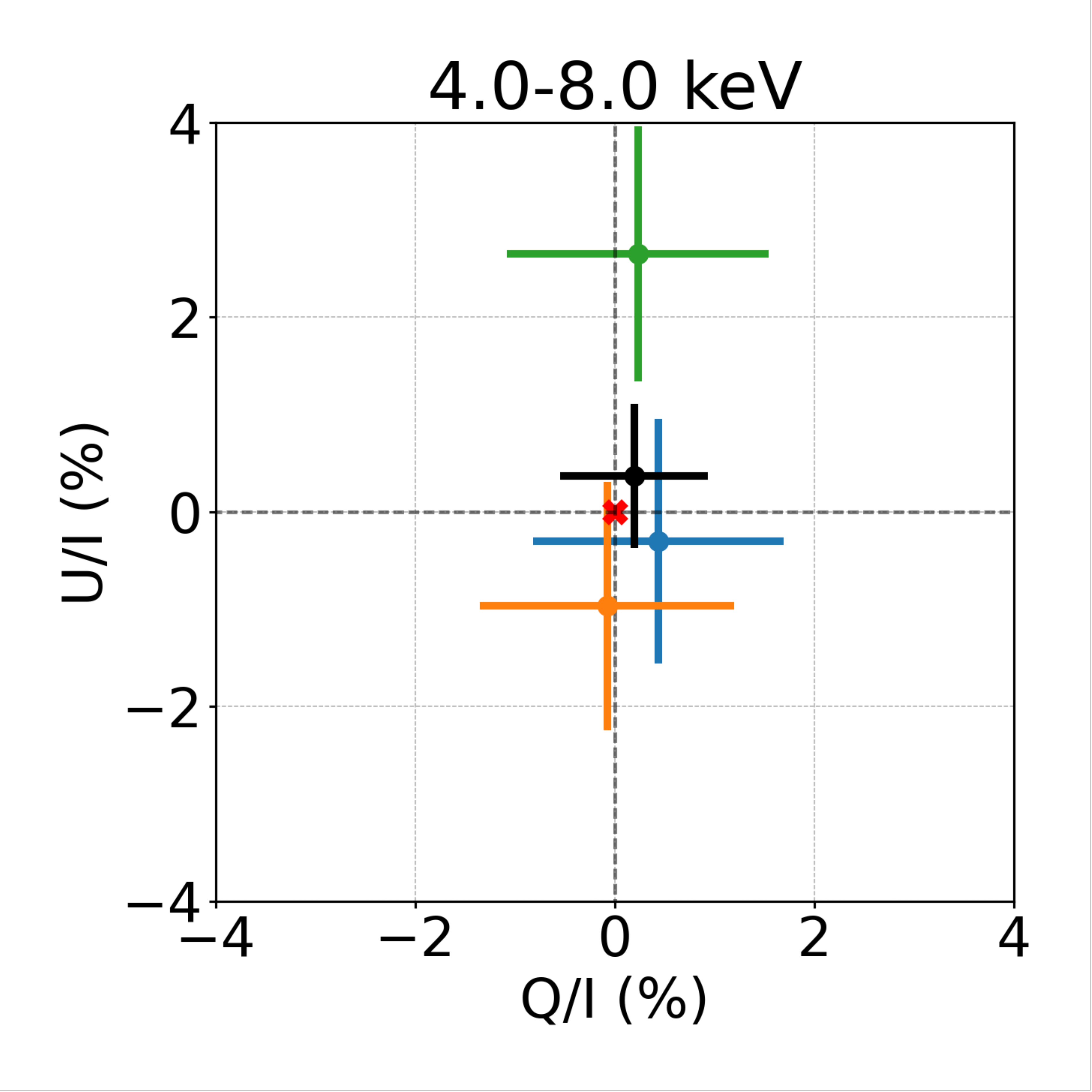}
\caption{Stokes parameters in the 2--8~keV, 2--4~keV and 4--8~keV energy band for DU1 (blue),  DU2 (orange) and DU3 (green) and by summing the events of the three DUs (black). The red point represents the null polarization.
}
\label{fig:pcube_stokes_2-4-8keV}
\end{figure*}

The Stokes parameters of \source observed by \ixpe in the 2--8, 2--4 and 4--8 keV energy bands, obtained 
with \textsc{ixpeobssim}, are reported in Table~\ref{tab:stokes} 
and in Figure~\ref{fig:pcube_stokes_2-4-8keV}. No detection of polarization can be claimed. We also analyzed the variation of Stokes parameters as a  function of time, but we did not obtain any significant detection.

\begin{deluxetable}{ccccc}
\tablecaption{Normalized Stokes parameters for  \source as observed by the three DUs of \ixpe in various energy bands \label{tab:stokes}.}
\tablecolumns{5}
\tablewidth{0pt}
\tablehead{
\colhead{ } &
\colhead{DU1} &
\colhead{DU2} &
\colhead{DU3} &
\colhead{All DUs}
}
\startdata
& \multicolumn{4}{c}{2--8 keV}  \\ 
$Q/I$    ~(\%)             & $\phantom{-}$0.48$\pm$0.63 & $\phantom{-}$0.14$\pm$0.65  & $-$0.11$\pm$0.66 & 0.18$\pm$0.37 \\
$U/I$   ~(\%)                      & $\phantom{-}$0.90$\pm$0.63 & $-$0.49$\pm$0.65 & $\phantom{-}$0.90$\pm$0.66 & 0.42$\pm$0.37 \\ \hline
& \multicolumn{4}{c}{2--4 keV} \\ 
$Q/I$    ~(\%)                & $\phantom{-}$0.50$\pm$0.62 & $\phantom{-}$0.23$\pm$0.63  & $-$0.26$\pm$0.65 & 0.17$\pm$0.37 \\
$U/I$    ~(\%)                     & $\phantom{-}$1.42$\pm$0.62 & $-$0.28$\pm$0.63 & $\phantom{-}$0.16$\pm$0.65 & 0.45$\pm$0.37 \\ \hline
& \multicolumn{4}{c}{4--8 keV} \\ 
$Q/I$   ~(\%)                 & $\phantom{-}$0.4$\pm$1.3 & $-$0.1$\pm$1.3  & $\phantom{-}$0.2$\pm$1.3 & 0.19$\pm$0.74 \\
$U/I$   ~(\%)                      & $-$0.3$\pm$1.3 & $-$1.0$\pm$1.3 & $\phantom{-}$2.7$\pm$1.3 & 0.37$\pm$0.74 \\ \hline
\enddata
\tablecomments{The values of the average modulation factors of the three DUs in various energy ranges are: 31.8\% (2--8 keV), 26.7\% (2--4 keV) and 43.6\% (4--8 keV), respectively. }
\end{deluxetable}

\begin{figure*}
\centering

\includegraphics[angle=0,scale=0.22]{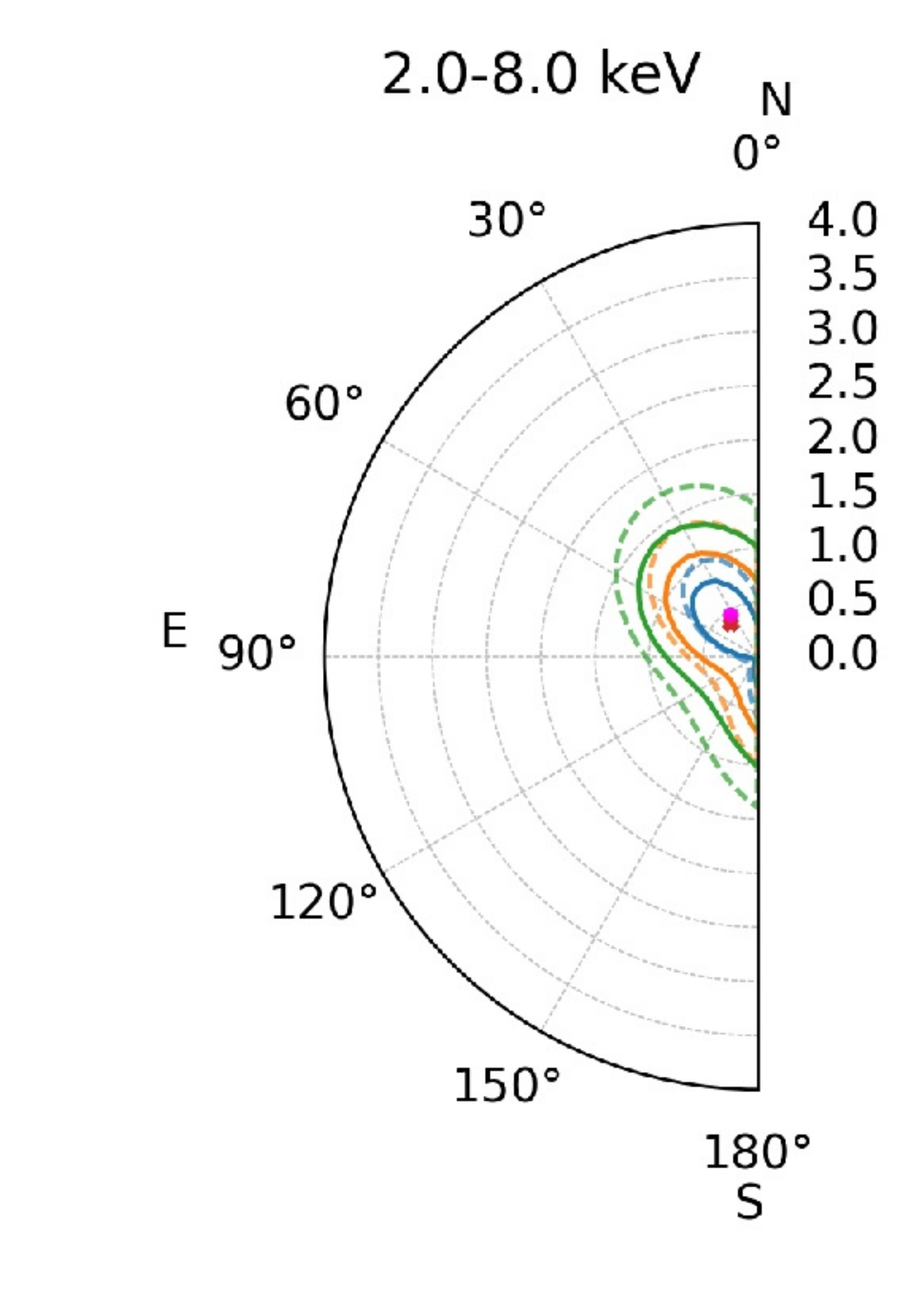}
\includegraphics[angle=0,scale=0.22]{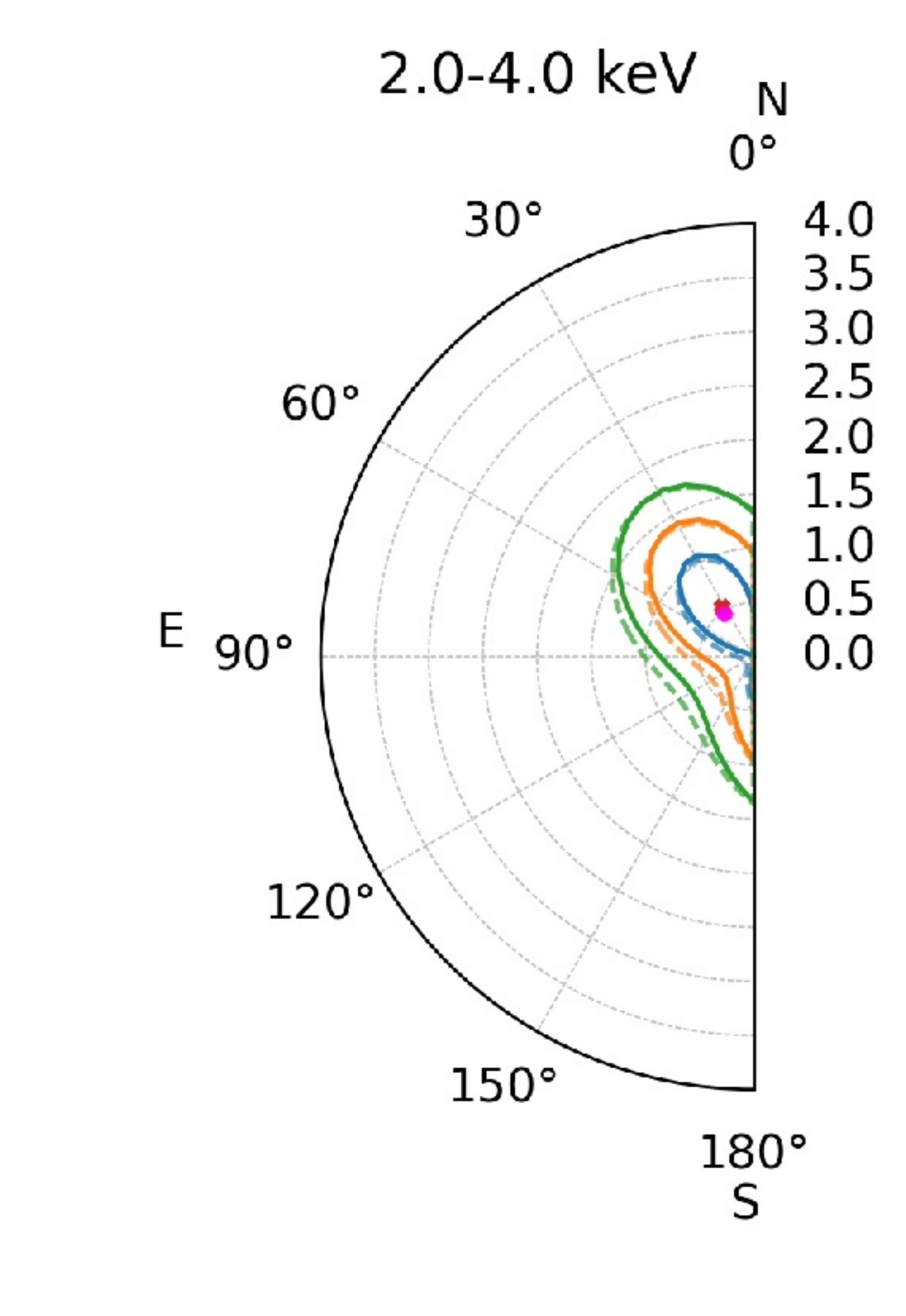}
\includegraphics[angle=0,scale=0.22]{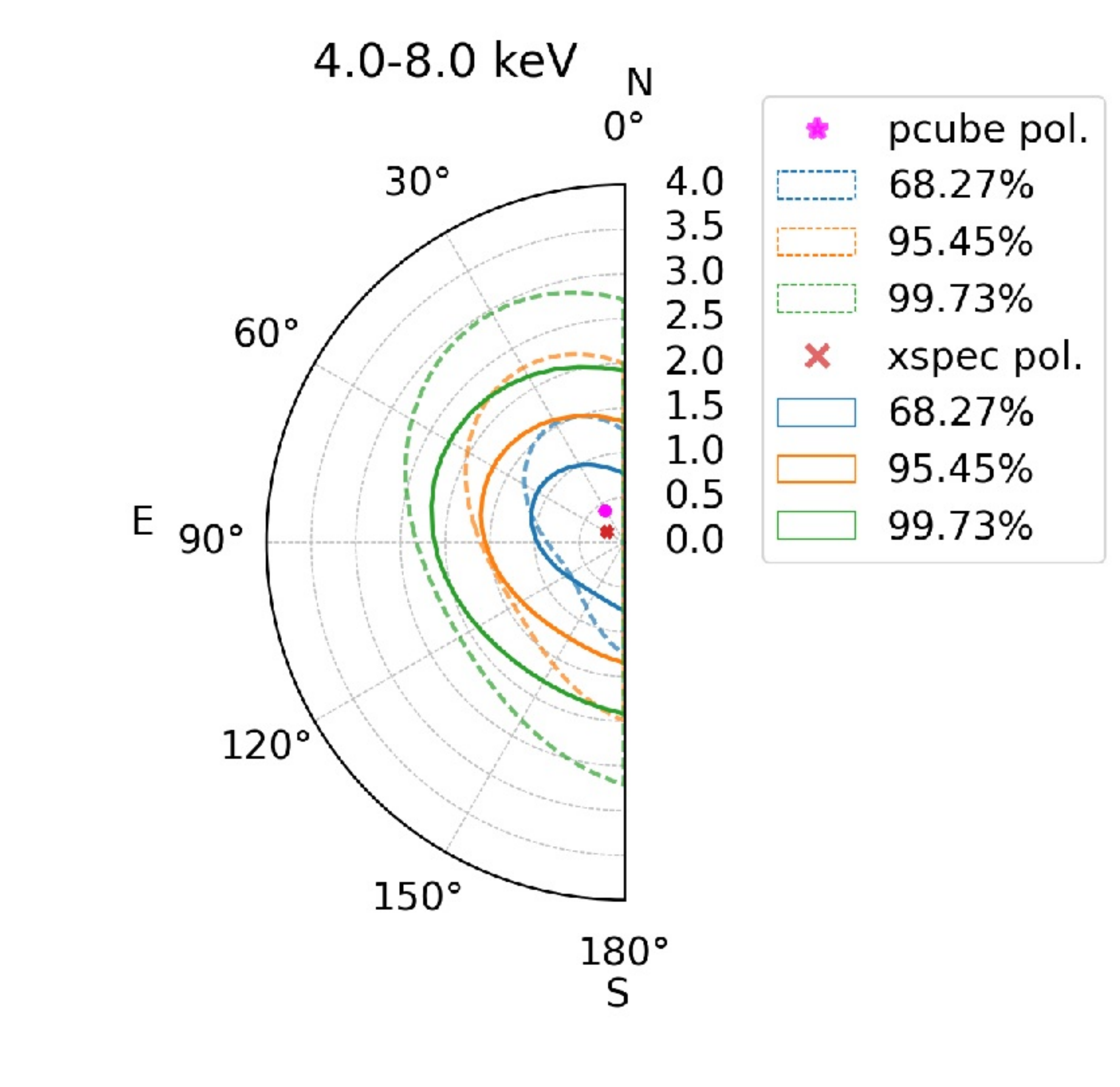}
\caption{Contour of the PD and PA for \source in the 2--4, 4--8 and  2--8~keV energy bands obtained with \textsc{xspec} (red cross and solid contours) and \textsc{ixpeobssim} (pink star and dashed contours) by summing the events from the three DUs. Contours correspond to the 68.27\%, 95.45\% and 99.73\% confidence levels.
 Both sets of contour levels, obtained with \textsc{xspec} and \textsc{ixpeobssim}, are computed for a joint measurement of PD and PA. Therefore, they are derived using a $\chi^2$ with two degrees of freedom.
}
\label{fig:pcube_frac_deg_2-4-8keV}
\end{figure*}
%

We performed the fitting spectro-polarimetric procedure by applying the {\sc polconst} convolution model to the \ixpe{} spectra ($I$, $Q$ and $U$) using \textsc{xspec} (syntax: {\polconst*\phabs(\diskbb+\comptt)}). This model describes a constant source polarization. In order to derive the polarization parameters (PD and PA of \polconst\ model, the spectral parameters of \phabs, \diskbb\ and \comptt\ models were fixed to those found from spectral fitting  of the \nicer and JEM-X data (see Table~\ref{tab:fitpar}).
As expected, the PD is compatible with null polarization and the PA is unconstrained even at a confidence level as low as 68.27\%. 
Table \ref{tab:polarization} reports the upper limits calculated with both \textsc{ixpeobssim} and \textsc{xspec} at different confidence levels.

Figure~\ref{fig:pcube_frac_deg_2-4-8keV} reports the contours of PD and PA of the \ixpe observation in the 2--8, 2--4 and 4--8~keV energy bands. They are obtained both with \textsc{xspec} (red cross and solid contours) and \textsc{ixpeobssim} (pink star and dashed contours) by summing the events from the three DUs. 
  The 1$\sigma$ upper limits on the PD from \textsc{ixpeobssim} (0.84, 0.85 and 0.94\% in the 2--8, 2--4 and 4--8 keV range, respectively, see Table~\ref{tab:polarization}) derived as described in \citet{Baldini2022} are somewhat larger than the estimates using a Bayesian approach presented by \citet{Maier}, which would give 0.56, 0.59, 0.82\%, but are consistent with the corresponding limits from the \textsc{xspec} of 0.69, 0.90, and 0.82\%. 
The \textsc{xspec} $3\sigma$ upper limits  (99.73$\%$ confidence level) are 1.3, 1.6, and 2.0\%, while the  
Bayesian approach gives rather consistent limits of 1.41, 1.44, and 2.37\%. 
In any case, the PA is unconstrained in all three energy bands (see Figure~\ref{fig:pcube_frac_deg_2-4-8keV}).

\begin{deluxetable}{ccc} 
\tablecaption{X-ray polarization of \source computed by means of \textsc{ixpeobssim} and \textsc{xspec}
\label{tab:polarization}}
\tablecolumns{3}
\tablewidth{0pt}
\tablehead{
\colhead{Energy Band} &
\colhead{\null }  &
\colhead{PD (\%)} 
}
\startdata
2--8 keV&  &  \\ \hline
&\textsc{ixpeobsim} @68.27$\%$ (1-$\sigma$) & $<$0.84\\
&\textsc{xspec} @68.27$\%$  &  $<$0.69\\
&\textsc{xspec} @99.73$\%$ & $<$1.3 \\
\hline
2--4 keV&  &  \\ \hline
&\textsc{ixpeobssim} @68.27$\%$ (1-$\sigma$)  &$<$0.85\\
&\textsc{xspec} @68.27$\%$  & $<$ 0.90 \\
&\textsc{xspec} @99.73$\%$  &  $<$1.6 \\ 
\hline
4--8 keV&  & \\ \hline
&\textsc{ixpeobssim} @68.27$\%$ (1-$\sigma$)   & $<$0.94\\
&\textsc{xspec} @68.27$\%$   & $<$0.82\\
&\textsc{xspec} @99.73$\%$ & $<$ 2.0\\
\enddata
\tablecomments{\textsc{ixpeobssim}  uncertainties are estimated assuming that variables are normally distributed, whereas \textsc{xspec} uncertainties are estimated by varying each parameter along $\chi^2$ surface. 
 The upper limits to the PD are obtained from the one-dimensional errors, without regard to the value of the PA. Thus, they are computed using a $\chi^2$ with one degree of freedom.
}
\end{deluxetable}

\section{Discussion and conclusions}

In order to put constraints on the geometry of the \source~ system, firstly we performed simulations  with the  general relativistic Monte Carlo code, {\sc monk}~\citep{Zhang19}, suitably adapted to compute the X-ray polarized radiation coming from weakly magnetized NS-LMXBs  in Kerr spacetime, accounting for the contributions of the neutron star, disk and corona \citep[see for details][and references therein]{Gnarini22}. 

As reported in \citet{Gnarini22}, a black-body spectrum is assumed to model the unpolarized neutron star surface emission, while the seed photons from the disk are generated according to the disk emissivity. 
The hot electron corona is illuminated by both the neutron star and the accretion disk and, when a photon reaches the corona, it is Compton scattered, assuming the Klein-Nishina cross section.
The energy and polarization spectrum is produced by counting the photons arrived to the observer.

\begin{figure*}[ht!] 
\centering
\includegraphics[angle=0,width=\textwidth]{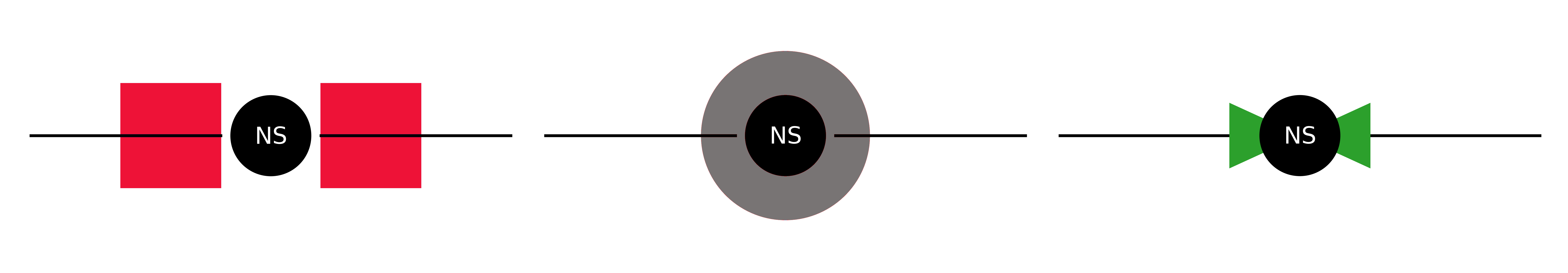}
\caption{Schematic representation of the three different geometries used in {\sc monk} simulations: the pseudo--toroidal geometry  (left panel), the shell (middle panel) and the wedge (right panel).}
\label{fig:geo}
\end{figure*}
\begin{figure*}[ht!]
\centering
\includegraphics[angle=0,scale=0.33]{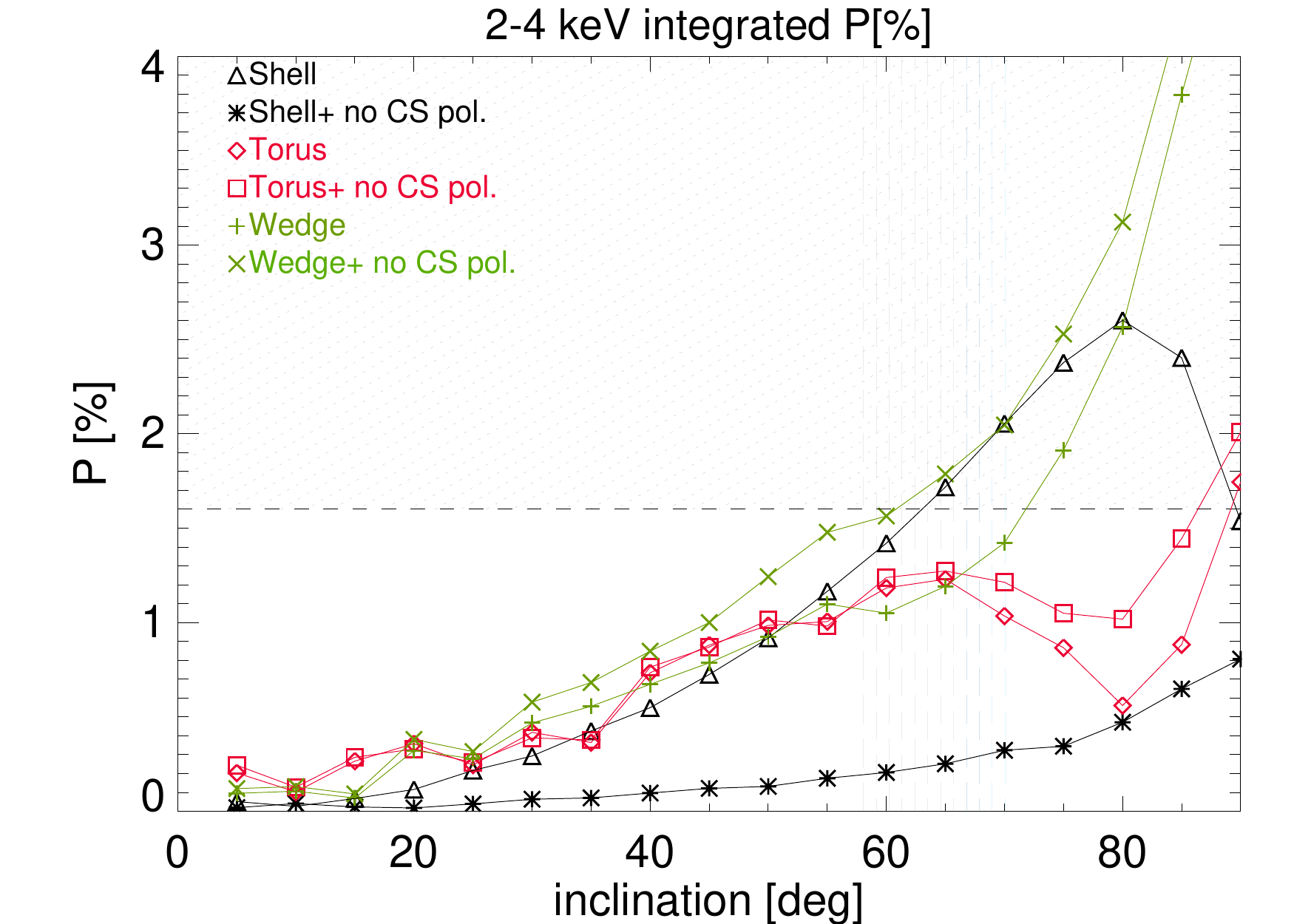}
\includegraphics[angle=0,scale=0.33]{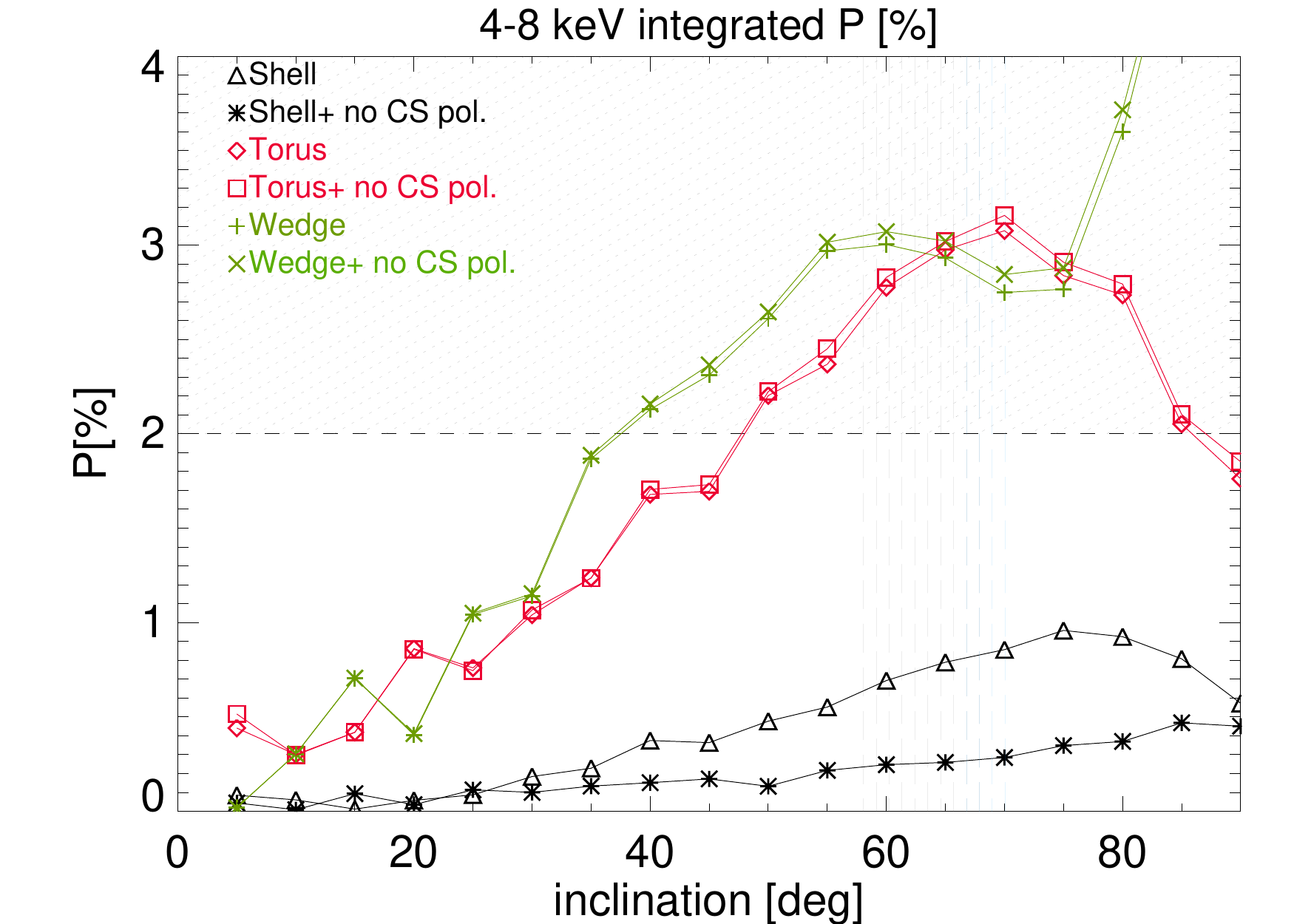}
\includegraphics[angle=0,scale=0.33]{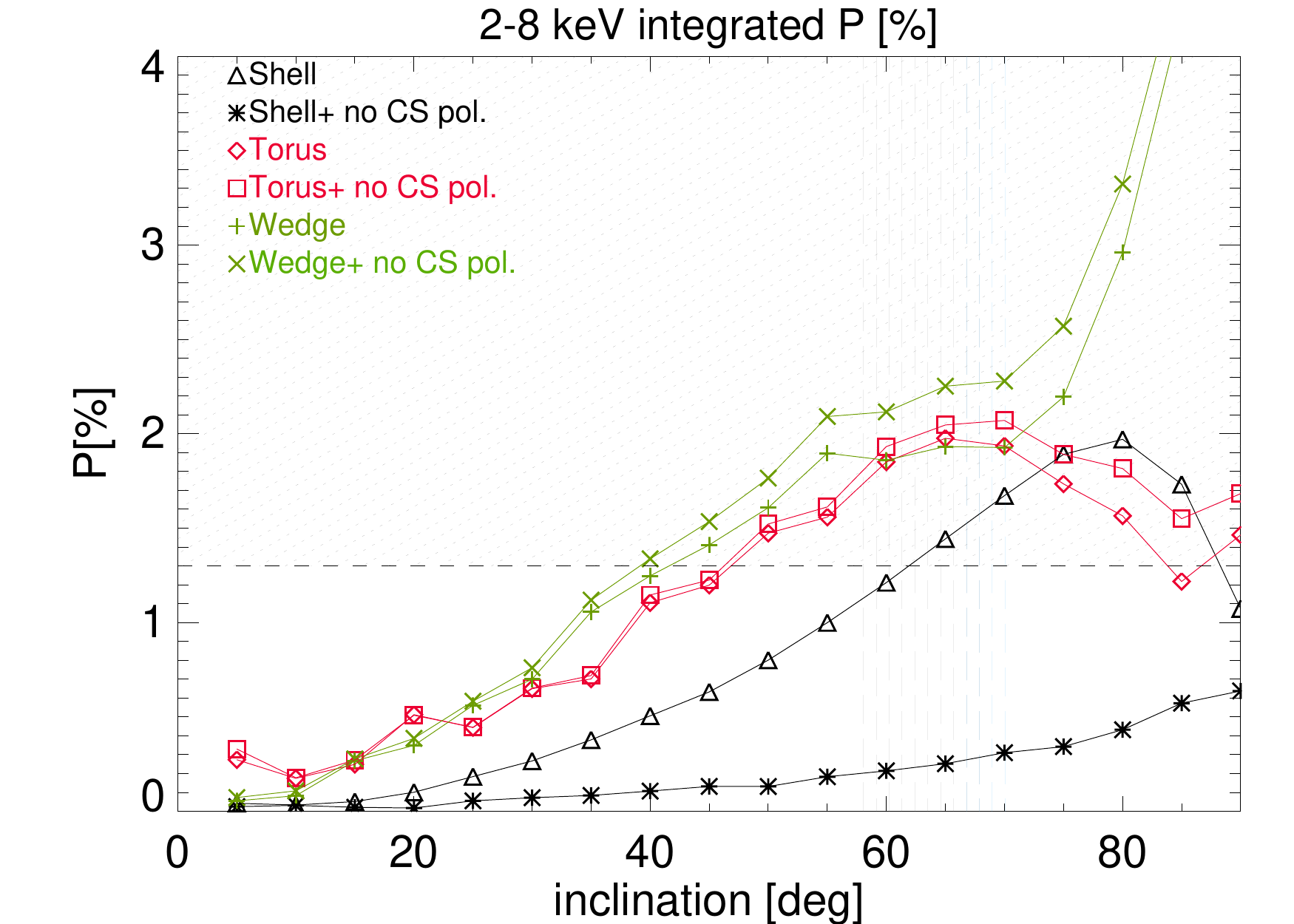}
\caption{Monk simulations of \source{} PD integrated over different energy ranges: 2--4 keV (left panel); 4--8 keV (middle panel); 2--8 keV (right panel) as a function of the  inclination angle. The black horizontal dashed lines represent the upper limits on the PD from Table~\ref{tab:polarization}. The upper gray hatched regions represent the values of PD excluded by our results. While the vertical dashed bands represent the interval of $i$ fixed by the indirect measurements of the \source inclination angle reported by both \citet{Johnston20},  black dashed lines, and \citet{Mescheryakov11},  light blue dashed lines. 
Results are given for two cases of disk seed photons: polarized according to the \citet{Chandrasekhar} law or unpolarized (labeled `no CS pol').} 
\label{fig:stokes_poldeg_simul}
\end{figure*}

The simulations were performed using as input parameters the  best-fit spectral parameters reported in Table~\ref{tab:fitpar} for different geometries and considering a standard neutron star with 1.4 $M_\Sun$, 12 km radius and 3 ms period, in analogy to the one derived from QPOs by \citet{Wijnands98} for Cygnus X-2 \citep[see also][for a statistical analysis of the spin distributions of NS-LMXBs]{Patruno17}.
In order to prove the presence, and eventually the geometry, of the electron corona and  to test if the nature of the hard component is instead strictly connected with the spreading layer, we performed simulations with three different geometries,  chosen among those implemented in the code, as shown in Figure~\ref{fig:geo}:

\begin{itemize}
\item Pseudo--toroidal geometry (as defined in ~\citealt{Gnarini22}): a rectangular section torus with similar vertical and horizontal length scales ($2H \sim\Delta R$)  corotating with the disk.  
As reported in \citet{Gnarini22}, the slab corona is assumed to cover only part of the disk, starting from the inner disk radius until 15 gravitational radii. While, the vertical thickness is set in order to cover most of the NS surface.

\item Shell geometry: a  stationary spherical shell surrounding the NS \citep[roughly mimicking the spreading layer of][]{Inogamov99}; we chose the same radius used in \citet{Gnarini22}.\footnote{Some preliminary tests on a co-rotating corona indicate that the PD is similar to the stationary case.} However, when varying the radius of the shell, the symmetry does not change, consequently the PD remains substantially unvaried.

\item Wedge geometry: a conical section torus around the NS equator lying between the disk and the NS surface and rotating with Keplerian velocity \citep[roughly mimicking the equatorial boundary layer, e.g.,][]{Popham01}. The torus is jointed to both NS surface and inner part of the disk (it extends from 6 to 8 gravitational radii).
\end{itemize}

Figure~\ref{fig:stokes_poldeg_simul} shows the net polarization fraction integrated over three different \ixpe{} energy bands as a function of the inclination angle, for the three different geometries. 
We also consider two cases for the polarization of the disk seed photons: polarized according to the \citet{Chandrasekhar} law for the semi-infinite, plane-parallel, pure electron-scattering atmosphere, and unpolarized (labeled `no CS pol' in Fig.~\ref{fig:stokes_poldeg_simul}).
The black dashed lines represent the \ixpe{} 3$\sigma$ upper limit for each energy band (see Table~\ref{tab:polarization}).
 
For pseudo--toroidal geometry, the presence of intrinsic polarization of disk seed photons does not significantly change the net fraction of polarized light since disk photons dominate only at lower energies. 
Therefore, we can derive a relatively stringent upper limit on the viewing angle:  $i \lesssim 47$\degr (see the right panel of Figure ~\ref{fig:stokes_poldeg_simul}). 
The PA for pseudo--toroidal geometry is misaligned and not perpendicular with respect to the disk, as results of the sum of disk and NS contributions together with GR effects \citep[see][for more details on this geometry]{Gnarini22}.
 
Considering the shell geometry, the presence or absence of intrinsic polarization could substantially change the PD while the PA is always parallel to the disk. 
In fact, for unpolarized disk seed photons, the fraction of polarized light remains well below 1\% for all inclinations in all the energy bands. On the other hand, in case of intrinsic polarization of the seed photons, a constraint on \source viewing angle is derived:  $i \lesssim 62$\degr (see left panel of Figure~\ref{fig:stokes_poldeg_simul}).

Finally, for the wedge geometry, the presence or the absence of intrinsic polarization slightly changes the polarization fraction giving upper limits of  $i \lesssim 42$\degr\ and $i \lesssim 39$\degr, respectively.  For this configuration, the PA is misaligned by approximately 25\degr\ from the projection of the rotation axis, as results of general and special relativity effects and the sum of the different photon populations.

On the other hand, comparing  the inclination values reported by previous authors, i.e. ${69^{+2}_{-3}}$~deg \citep{Johnston20}, and $62\fdg5 \pm5\fdg5$ \citep{Mescheryakov11}, with our simulations, both pseudo–toroidal and wedge
geometries seems to be excluded. 
In fact as the plots in Figure~\ref{fig:stokes_poldeg_simul} show, for inclinations between 57$\degr$ and 72$\degr$ there should be a detection of polarization at least in one of the three considered energy ranges (see the dashed rectangle in the three panels of Figure~\ref{fig:stokes_poldeg_simul}).  For the shell geometry and no intrinsic polarization, there is no detection within the interval of viewing angles considered in all the three energy ranges. On the contrary, in the case of shell geometry and intrinsic polarization, the interval of viewing angles with no detection is restricted to $57\degr\lesssim i \lesssim 62\degr$  excluding the values of inclination reported by \citet{Johnston20}, $i\sim69^{+2}_{-3}$~deg, but not those reported by \citet{Mescheryakov11}, $i\sim 62\fdg5 \pm5\fdg5$.
However, the results are computed using only the value of best-fit parameters without including the errors. These can lead to slight variations on the inclination constraints.
Therefore, either the \source system could have a spherical symmetry or its inclination is lower than previously measured.  In fact, most of the simulations show \citep[see, e.g.,][]{Gnarini22,Schnittman} that small viewing angles correspond to a lower fraction of polarized light emitted by a source. 

We have to underline that a significant percentage of polarized light was measured in various LMXBs, such as, for example, the mentioned Sco~X-1 and recently Cyg X-2 (Farinelli et al., submitted). Both sources are observed at inclination angles comparable with that of \source. However, these two sources are classified as Z sources, while  \source is the first atoll source observed by \ixpe. A comparison between the two kind of sources is not always possible. For example, \citet{Long} report that in Sco X-1 the PD has a strong dependence on the luminosity and the spectral branch. Instead, the \ixpe data of \source present  quite stable light curve and hardness ratio. Therefore, it is impossible to extract any information about the evolution of the PD as a function of luminosity and the spectral state unlike the case of Sco X-1.

As reported by   \citet{Lapidus85} and \citet{Schnittman}  the reflection from the accretion disk  of the radiation produced by the SL or self-illumination of the disk can produce substantial polarization.
However, we do not detect in  \source, at least with the spectral resolution of \nicer, the iron line that is a typical signature of disk reflection in the HSS sources \citep[e.g. Cyg X-2 and Sco X-1;][]{Dai,DiSalvo}. 
One possibility is that the disk is strongly ionized reducing the strength of the iron line. 
On the other hand, the latitudinal extent of the SL might not be large enough to produce significant illumination of the disk resulting in a weak signal. This could be one of the reasons why we could only establish an upper limit for polarization in \source.

Finally, by significantly varying the dimension of the hot corona or considering more complicated shapes (e.g. a combination of two proposed geometries), the PD could be very different compared to the previous cases. However, if we assume that the spherical geometry, that seems favored by the simulations, mimic the SL (thus the SL subsume the role of the corona), it  could not be extended more than some fraction of the NS radius (the same line of thinking could be applied in case of the boundary layer). 
Furthermore, two different emission components (for example, the disk and the reflection component) or two different populations of electrons emitting in different regions, may have similar PD but orthogonal PA.

\begin{acknowledgments}
The {\it Imaging X-ray Polarimetry Explorer} (\ixpe) is a joint US and Italian mission.  The US contribution is supported by the National Aeronautics and Space Administration (NASA) and led and managed by its Marshall Space Flight Center (MSFC), with industry partner Ball Aerospace (contract NNM15AA18C).  The Italian contribution is supported by the Italian Space Agency (Agenzia Spaziale Italiana, ASI) through contract ASI-OHBI-2017-12-I.0, agreements ASI-INAF-2017-12-H0 and ASI-INFN-2017.13-H0, and its Space Science Data Center (SSDC) with agreements ASI-INAF-2022-14-HH.0 and ASI-INFN 2021-43-HH.0, and by the Istituto Nazionale di Astrofisica (INAF) and the Istituto Nazionale di Fisica Nucleare (INFN) in Italy.  This research used data products provided by the IXPE Team (MSFC, SSDC, INAF, and INFN) and distributed with additional software tools by the High-Energy Astrophysics Science Archive Research Center (HEASARC), at NASA Goddard Space Flight Center (GSFC).

This research used data products provided by the \ixpe Team (MSFC, SSDC, INAF, and INFN) and distributed with additional software tools by the High-Energy Astrophysics Science Archive Research Center (HEASARC), at NASA Goddard Space Flight Center (GSFC). 
This work made use of the MAXI light curves, publicly available at http://maxi.riken.jp/top/slist.html.
INTEGRAL is an ESA project with instruments and science data centre funded by ESA member states (especially the PI countries: Denmark, France, Germany, Italy, Switzerland, Spain) and with the participation of Russia and the USA. We thanks Keith Gendreau, Craig Markwardt and the NICER SOC for granting and performing the NICER observations of the source, and helping with data reduction.

JP and SST acknowledge support from the Russian Science Foundation grant 20-12-00364 and the Academy of Finland travels grants 349144, 349373, and 349906. 
JP and JJEK are supported by the Academy of Finland  grant 333112.  
\end{acknowledgments}

\vspace{20mm}
\facilities{IXPE, Swift (XRT and UVOT), INTEGRAL}

\software{Stingray \citep{Stingray1,Stingray2},  
          \textsc{xspec} \citep{Arnaud1996}, 
          \textsc{ixpeobssim} \citep{Baldini2022},
          \textsc{monk} \citep{Zhang19}.
          }
          



\bibliography{ixpe_ns}
\bibliographystyle{aasjournal}



  \end{document}